# Unveiling the interaction mechanisms of electron and X-ray radiation with halide perovskite semiconductors using scanning nano-probe diffraction

Jordi Ferrer Orri,[1,2] Tiarnan A.S. Doherty,[1] Duncan Johnstone,[2] Sean M. Collins,[3] Hugh Simons,[4] Paul A. Midgley,[2] Caterina Ducati,[2,*] and Samuel D. Stranks[1,5,*]

[1]Cavendish Laboratory, University of Cambridge, Cambridge, UK.
[2]Department of Materials Science and Metallurgy, University of Cambridge, Cambridge, UK.
[3]School of Chemical and Process Engineering & School of Chemistry, University of Leeds, Leeds, UK.
[4]Department of Physics, Technical University of Denmark, Copenhagen, Denmark.
[5]Department of Chemical Engineering and Biotechnology, University of Cambridge, Cambridge, UK.
*Corresponding authors: cd251@cam.ac.uk, sds65@cam.ac.uk

**The interaction of high-energy electrons and X-ray photons with soft semiconductors such as halide perovskites is essential for the characterisation and understanding of these optoelectronic materials. Using nano-probe diffraction techniques, which can investigate physical properties on the nanoscale, we perform studies of the interaction of electron and X-ray radiation with state-of-the-art $(FA_{0.79}MA_{0.16}Cs_{0.05})Pb(I_{0.83}Br_{0.17})_3$ hybrid halide perovskite films (FA, formamidinium; MA, methylammonium). We track the changes in the local crystal structure as a function of fluence using scanning electron diffraction and synchrotron nano X-ray diffraction techniques. We identify perovskite grains from which additional reflections, corresponding to $PbBr_2$, appear as a crystalline degradation phase after fluences of ~200 $e^-Å^{-2}$. These changes are concomitant with the formation of small $PbI_2$ crystallites at the adjacent high-angle grain boundaries, with the formation of pinholes, and with a phase transition from tetragonal to cubic. A similar degradation pathway is caused by photon irradiation in nano-X-ray diffraction, suggesting common underlying mechanisms. Our approach explores the radiation limits of these materials and provides a description of the degradation pathways on the nanoscale. Addressing high-angle grain boundaries will be critical for the further improvement of halide polycrystalline film stability, especially for applications vulnerable to high-energy radiation such as space photovoltaics.**

# 1. Introduction

Halide perovskite materials exhibit promising characteristics for optoelectronics applications. These materials can be fabricated through facile processing techniques and yield high performance devices, albeit with large compositional and structural heterogeneities at multiple length scales.[1] Heterogeneity, particularly on the nano- and microscale, is key to understanding some of the critical questions on performance and stability.[1–3] Scanning nanoprobe characterisation techniques, such as scanning electron diffraction (SED) or nanoprobe X-ray diffraction (nXRD), are particularly advantageous as they can access both the relevant nano- and microscales.[4,5]

Both techniques use focused radiation, which interacts with the material and can impact its structure and chemistry, especially for materials with complex stoichiometries such as hybrid halide perovskites that mix organic and inorganic ions. These hybrid halide perovskites are uniquely promising light-weight candidates for space photovoltaic applications, with reports showing excellent radiation hardness under extreme space conditions of proton and electron exposure.[6–8] Large fluxes of high-energy particles, such as trapped electrons in orbit extending up to 1 MeV energies, can trigger defect generation and degradation of these materials, in similar ways to the degradation induced during characterisation techniques.[9,10] Based on studies reporting radiation damage of hybrid perovskite being dependent on the total dose rather than on the dose rate,[11,12] microscopy techniques such as SED and nXRD may be used to mimic large total radiation exposures whilst characterizing defect generation and degradation in vacuum.

Understanding the interaction of the high-energy probe with the samples is critical for acquisition, data analysis, and detector applications. For example, the interaction of focused electron beams with halide perovskites has been reviewed and can be quantified with critical radiation values. These critical values are defined as particle fluences (in units of particles $Å^{-2}$) representing the rate of characteristic irreversible structural changes under particle exposure.[12–14] For halide perovskites, for example, higher acceleration voltages increase these critical values by reducing the cross-section for beam-specimen interaction, while the use of cryogenic temperatures induces rapid amorphisation and reduces these values.[11,14] Rothmann *et al.* used transmission electron microscopy (TEM) to quantify structural changes in methylammonium lead iodide films ($MAPbI_3$) at fluences as low as 100 e⁻ $Å^{-2}$,[11] in which loss of the organic moieties results in lattice contraction and the formation of a supercell, further described by Chen *et al.* as $MAPbI_{2.5}$, ultimately degrading into $PbI_2$.[15,16] The appearance of an intermediate phase of degradation agrees with similar studies by scanning electron microscopy techniques.[17–19] Alberti *et al.* also reported the detrimental effect of having excess Pb-related defects during fabrication, which aggregate and feed degradation at grain boundaries upon electron irradiation.[20] However, $Pb^0$ formation is usually observed as a degradation product of fully-inorganic compositions, such as $CsPbI_3$, upon electron irradiation. This is because the inorganic cation is less prone to be reduced than its organic counterparts or than the Pb cation.[21,22]

Moving beyond the workhorse hybrid $MAPbI_3$ composition, Rothmann *et al.* recently reported high-resolution scanning transmission electron microscopy (STEM) images of evaporated formamidinium lead iodide ($FAPbI_3$) thin films, reporting the appearance of additional reflections at low fluences of 200 e$^-$ Å$^{-2}$, the loss of the perovskite phase at 600 e$^-$ Å$^{-2}$, and the eventual formation of $PbI_2$ grown on the lattice-expanded degraded phase of $FAPbI_3$ after 1,000 e$^-$ Å$^{-2}$.[23]

The effect of scanning X-ray probes on halide perovskites, on the other hand, has been less studied than for electron microscopy.[5] The critical fluences for $MAPbI_3$/$MAPbBr_3$ derivatives vary for different microscopy techniques. The nano X-ray beam-induced current signal is highly sensitive to defects and was rapidly affected at the lowest fluences of $10^3$ photons Å$^{-2}$.[24] Nano X-ray diffraction (nXRD) and fluorescence (nXRF) were found to withstand 10x higher exposures, since these signals are more intense, generated by larger volumes, and easier to record.[25,26] Li *et al.* reported similar critical fluences of $10^2$ and $10^4$ photons Å$^{-2}$ for $MAPbBr_3$ and $CsPbBr_3$ single crystals, respectively, yet the appearance of additional phases was not reported at these fluences.[25] Other studies have reported comparable XRD signal decays due to loss of crystallinity after long X-ray exposures.[27] These studies, specifically identifying the effect of synchrotron scanning probes on the material, are not to be confused with the plethora of laboratory XRD reports elucidating the degradation products of halide perovskites at millimeter scale due to external factors such as light, air, temperature or passivating agents.[28–30]

While the majority of the work related to beam damage has been performed on $MAPbI_3$, state-of-the-art halide perovskites are FA-rich mixed-cation mixed-halide compositions, especially those relevant for tandem and space photovoltaic applications.[31,32] Such compositions exhibit distinct chemical and structural degradation pathways, yet no comprehensive studies on such systems have been reported to date. Rigorously establishing the radiation effects on these more complex compositions to allow elucidation of global mechanisms, as well as obtaining insight into the role of grain boundaries and complicated junctions, is of immense interest.

Here we use two advanced nanoprobe diffraction techniques, SED and nXRD, to understand the respective interaction of 200 keV electrons and 20 keV X-rays with the local nanostructure of solution-processed halide perovskite films of triple-cation composition $(FA_{0.79}MA_{0.16}Cs_{0.05})Pb(I_{0.83}Br_{0.17})_3$.[33] By using low-dose diffraction microscopy, we can spatially resolve the diffraction information on the nanoscale as a function of radiation exposure. The use of finely tunable acquisition conditions allows us to track the compositional and structural evolution of degradation processes of hybrid perovskite films, as well as their spatial origin. We show that change is predominantly observed at specific sites in the polycrystalline film which are more defective and evolve towards new structures over time, such as $PbI_2$ and $PbBr_2$ formation. Specifically, high-angle grain boundaries in the polycrystalline structure trigger such changes to the nanostructure. These studies establish critical radiation values and interaction mechanisms of electron and X-ray radiation for mixed-cation mixed-halide perovskites, allowing the elucidation of global mechanisms for degradation, as well as stability windows in both measurement and application of these radiation types.

## 2. Results and Discussion

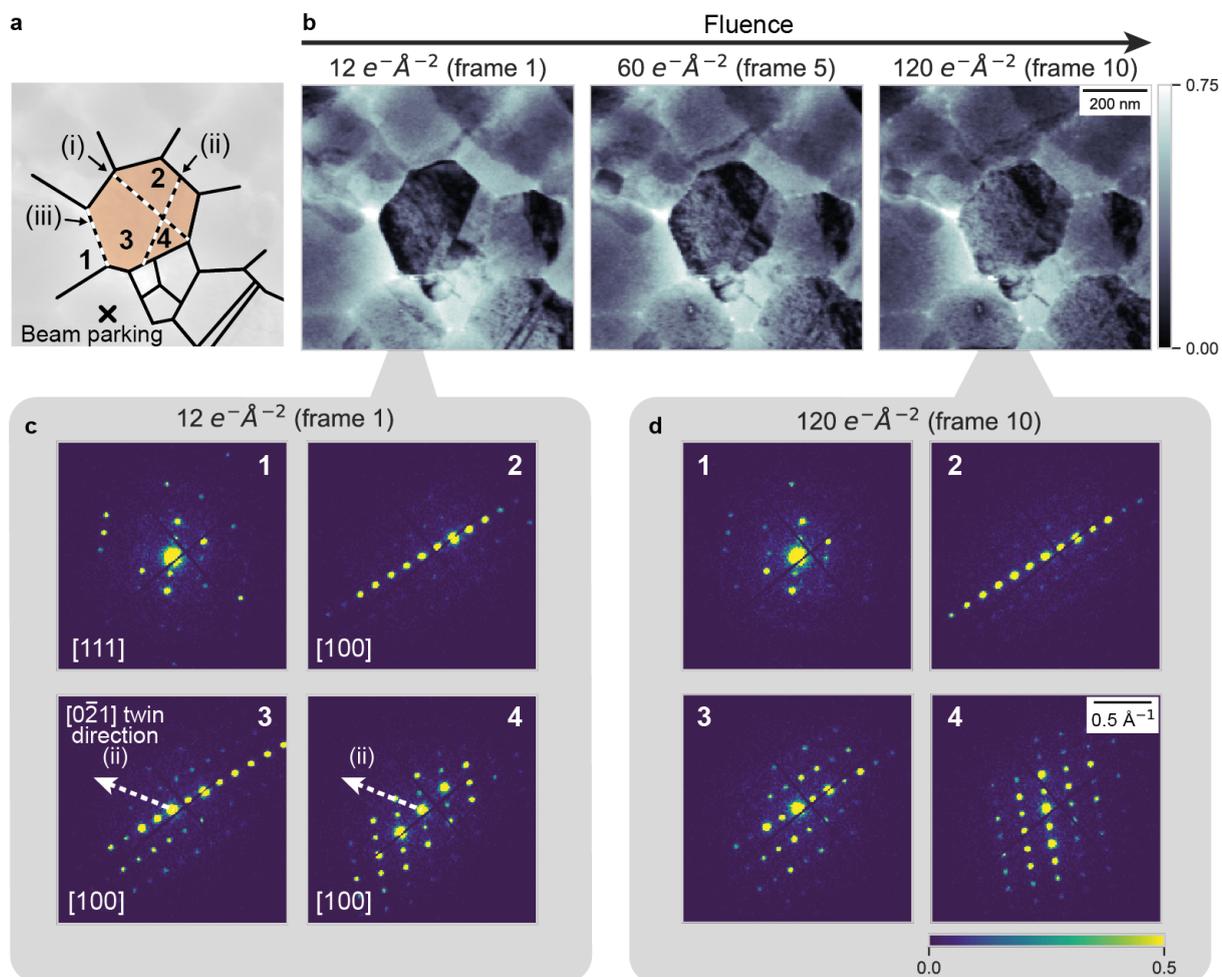

**Figure 1: Visualizing grain boundary types and their evolution under low electron exposure using scanning electron diffraction (SED) on $(FA_{0.79}MA_{0.16}Cs_{0.05})Pb(I_{0.83}Br_{0.17})_3$ halide perovskite films.** a) Schematic of the perovskite grain showing the areas of interest from which diffraction patterns are extracted (in Arabic numerals), and three interfaces between crystallographic domains indicated by dashed lines (in Greek numerals): (i) grain bending across region 2 and 3, (ii) a twin grain boundary at $(0\bar{2}1)$ plane between region 3 and 4, and (iii) high-angle grain boundary interphasing two grains at different zone axes between region 1 and 3. The beam-parking position is labeled with a cross. b) Evolution of vBF images of a perovskite film for up to 120 e$^-$ Å$^{-2}$ fluence. c) The resulting averaged diffraction pattern for four different areas at the 1$^{st}$ frame (12 e$^-$ Å$^{-2}$) and d) at the 10$^{th}$ frame (120 e$^-$ Å$^{-2}$). Regions denoted with numerals in (a) correspond to the diffraction in (b) and (c). All intensity scale bars correspond to the normalized vBF and the diffraction intensity.

Thin films of $(FA_{0.79}MA_{0.16}Cs_{0.05})Pb(I_{0.83}Br_{0.17})_3$ were solution-processed following optimized protocols for device fabrication compatible with silicon nitride (SiN) grids, ensuring reproducible and comparable results to high-performance devices prepared on standard substrates,[34] while also enabling the use of transmission microscopy techniques. 5-dimensional datasets of the films were acquired by combining a time series (t) of SED frames. Each SED frame was acquired over the same region of the sample (x,y), wherein each scanned position contains an electron diffraction pattern ($k_x$, $k_y$). Between each frame, the electron beam was digitally driven to a controlled beam-parking position within the field of view before being blanked, labeled with a cross in **Figure 1**. Visible beam-induced accelerated changes are observed in the vicinity of the beam-parking position, within a radius of approximately ~90 nm, equivalent to the radius of the interaction volume (see Figure S1). The circular area around the beam parking position is disregarded, using the rest of the scan area for this study, only irradiated during beam rastering.

The rich diffraction-based multidimensional dataset enables the investigation of the effects of cumulative electron beam exposure on the nanoscale. Repeated scans were acquired at total fluences from 12 $e^- Å^{-2}$ (1 frame) up to 900 $e^- Å^{-2}$ (75 frames), at which point the diffraction from the halide perovskite film had substantially degraded. Figure 1 shows the evolution of the local nanoscale diffraction of a representative perovskite grain at low electron exposure, up to a cumulative fluence of 120 $e^- Å^{-2}$ corresponding to the first 10 frames. The term grain is used here as the apparent area delimited by morphological grain boundary features observed in the electron microscope.[35,36] For each diffraction pattern, electrons at low scattering angle (the direct beam) are collected at the origin, which is surrounded by a series of diffraction peaks corresponding to scattering vectors near or at the Bragg condition. We create virtual bright-field images (vBF) of the scanned region by mapping the intensity of the direct beam below a virtual aperture of radius $4 \cdot 10^{-3}$ $Å^{-1}$ (semi-angle of detector of 2.39 mrad), as a function of the probe position. Virtual dark-field images (vDF) can be created by mapping the intensity of selected scattered signals, corresponding to a fixed virtual aperture size and position in reciprocal space within 1.18 $Å^{-1}$ range (70.3 mrad, see Figure S2). The vBF images mostly contain information from the non-scattered beam and low-angle inelastic scattering, resulting in relatively lower intensities when the sample is at a strongly diffracting orientation or where it is thicker.

The grain labeled in Figure 1a exhibits intense diffraction as their orientations match a low-angle zone axis, and is thus selected for detailed study. The vBF images in Figure 1b reveal the polycrystalline nature of the perovskite film, as well as heterogeneity in the diffraction contrast within each grain. These variations in contrast are a result of chemical and structural heterogeneity on the nanoscale.[1,37] To illustrate the difference between individual grains, diffraction patterns are obtained by summing pixels within the grain at the center of the region of interest, denoted by Arabic numerals in Figure 1a. The diffraction patterns at the near-pristine state are shown in Figure 1c, acquired at the lowest electron fluence of 12 e⁻ Å⁻² (first frame).

All diffraction patterns can be indexed to a tetragonal perovskite structure in the P4/mbm space group with lattice parameters of $a = b = 9.00$ and $c = 6.36$ Å,[38] each near the [100] zone axis with slightly different crystal tilt relative to the electron beam (see Figure S3-5 for simulated diffraction pattern matching experimental diffraction). At least three different crystal orientations can be observed within this grain in this scattering geometry. The interfaces between crystallographic domains can be identified in the virtual images and labeled by dashed lines and Greek numerals in the schematic (also see Figure S6). Grain bending due to the soft perovskite lattice is observed in (i), across the upper and the lower parts of the central grain in Figure 1b (region 2 and 3, respectively). The difference between their diffraction patterns can be attributed to small in-plane crystal tilt of ~2° (see Figure S7). When sudden changes in crystal orientation are observed, different types of grain boundaries can be identified.[39] In (ii) a twin boundary, a special case of a large angle grain boundary for which there is no atomic misfit, is observed between the regions 3 and 4 in Figure 1b. Detailed analysis of the diffraction pattern shows the lattice being mirrored across the $(0\bar{2}1)$ plane, marked with a dashed line. \distinctly, the grain boundaries around the central grain are (iii) high-angle, as inferred from orientation analysis on the adjacent grains being near the [111] zone axis instead (see Figure 1b-c region 1, for the diffraction pattern of an adjacent grain). The nature of these two types of grain boundary is discussed in more detail later.

The evolution of the diffraction patterns in Figure 1, after 120 e⁻ Å⁻² of accumulated fluence (10 frames), reveals that the strong diffraction contrast seen in the vBF images rapidly becomes more homogeneous. The diffraction patterns at the 1st and 10th frames show a small relative tilt of the order of a few degrees towards alignment to the [100] zone axis. We attribute these tilts to microstructural changes on the nanoscale, noting that bending of the support membrane cannot be excluded. This tilting is visible in the regions labeled as 3 and 4 in Figure 1c, showing more evenly distributed diffraction spots near the center of the beam after 10 scans. These tilts are not observed in the region labeled as 2, where the diffraction pattern, originally off the zone axis by a larger tilt, cannot rotate towards the zone axis. These observations suggest that changes in the perovskite microstructure may already start at extremely low electron exposure of tens of e$^-$ Å$^{-2}$, sufficient to provide enough energy to modify the pristine grain orientation (see Figure S7 for a more detailed evolution at low fluences of these diffraction patterns). Similar changes in the tilt of grains, commonly seen in thin polycrystalline films deposited on TEM substrates, are observed from the diffraction patterns in most of the grains in the scanned region, as well as in other SED experiments taken from other halide perovskite films of the same composition (see Figure S8).

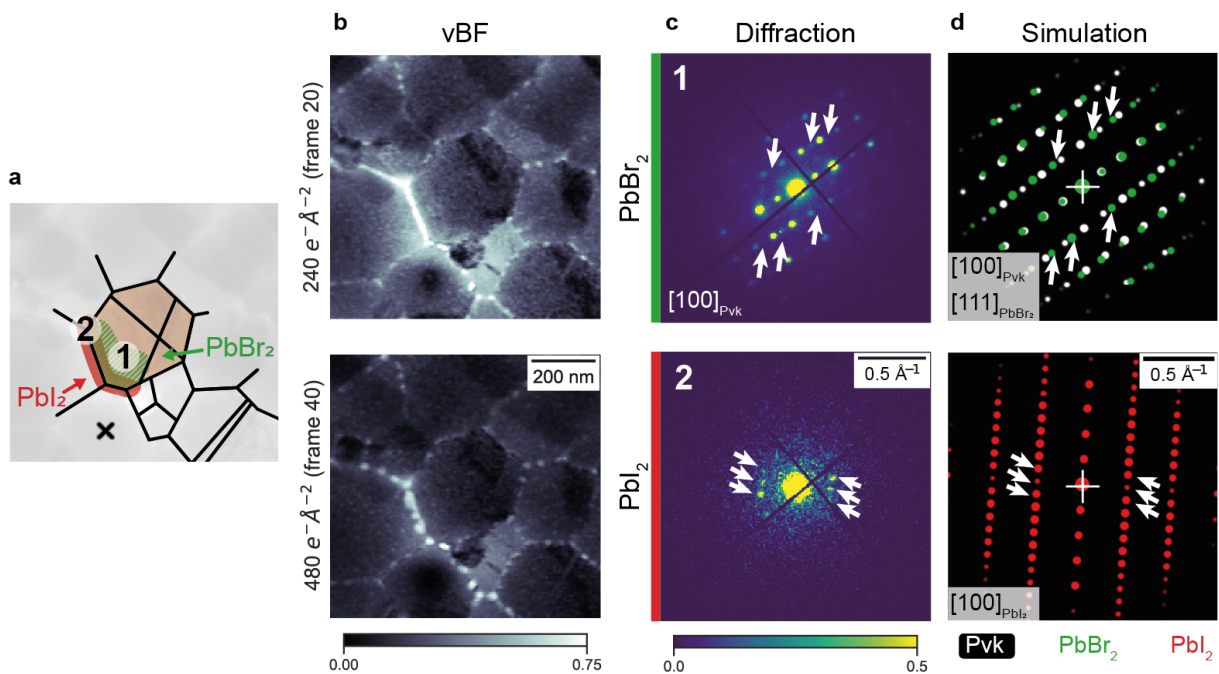

**Figure 2: Probing the emergence of lead halide species on the nanostructure of (FA$_{0.79}$MA$_{0.16}$Cs$_{0.05}$)Pb(I$_{0.83}$Br$_{0.17}$)$_3$ halide perovskite films at high electron exposure using SED.** a) Schematic of the perovskite grain showing the areas of interest from which PbI$_2$ and PbBr$_2$ species is observed and from which diffraction patterns are extracted, labeled as 1 and 2, respectively. b)The vBF images at the 20$^{th}$ and 40$^{th}$ frame. c) For each frame, a normalized diffraction pattern from the labeled region is shown. Some additional diffraction spots, marked with white arrows, are observed: the extra reflections in the region 1 can be indexed to PbBr$_2$ near the [111] zone axis, and the extra reflections in the region 2 can be indexed to PbI$_2$ near the [100] zone axis. d) Simulated patterns for the perovskite (Pvk), the PbBr$_2$ and the PbI$_2$ phases in white, green and red, respectively. The diffraction patterns computed using CrystalMaker match those seen experimentally.

We now consider further electron exposure by taking cumulative fluences up to 240 e$^-$ Å$^{-2}$, as shown in **Figure 2a**. The evolution of vBF contrast beyond 240 e$^-$ Å$^{-2}$ fluence suggests the formation of pinholes at the high-angle grain boundaries (Figure 2b). Degradation at the grain boundaries can be attributed to the loss of chemical species and lattice contraction due to the transformation from pristine perovskite to lead halide degradation products (see Figure S9 for the histograms of the Pb-Pb distances for the perovskite structure, and some of the lead halide structures with shorter Pb-Pb distances). Similar observations have been reported in TEM mode.[11] Overall, there is a constant decrease in the total intensity of the vBF images by ~25% that occurs uniformly across the whole scanned area over 240 e$^-$ Å$^{-2}$ (20 frames), seen in Figure 1b. Further electron exposure results in roughening of the grain and a decrease of the total scattering intensity by ~50% after 480 e$^-$ Å$^{-2}$ (40 scans), seen in Figure 2b. We note that such a large change in the vBF intensity cannot be attributed to fluctuations of the electron emission gun alone (Figure S10). The loss of intensity of the direct beam could be attributed to the sample thickening by deposition of an amorphous carbon layer on the surface (see the supplementary information, SI), and to possible densification of the perovskite structure by amorphisation or the loss of the lighter elements such as the organic cations. These changes would increase scattering, as seen in the virtual images created from the annularly integrated vDF (Figure S10).

We now inspect how the diffraction patterns near the [100] zone axis, taken from the central grain, change in Figure 2c. Weak Bragg reflections emerge from region 1 in the grain in Figure 2a after 240 e$^-$ Å$^{-2}$ (20 frames). These additional reflections, marked with white arrows, cannot be indexed to the pristine perovskite structure at $k$ = 0.23, 0.34 and 0.47 Å$^{-1}$. Close assessment reveals that the patterns from this region are indexable to PbBr$_2$ at the [111] zone axis within a 0.01 Å$^{-1}$ error, as shown in Figure 2d. This suggests the presence of these lead halide species epitaxially growing on the perovskite grain. We ruled out the possibility of this being due to superlattice reflections from the orthorhombic Pnma space group structure at the [001] zone axis, since the reflections are not located at the expected ½($hkl$) positions (see indexation in Figure S3).[15,40]

In contrast, a diffraction pattern taken from region 2 in Figure 2a after 480 e$^-$ Å$^{-2}$ exposure (40 frames) shows diffuse diffraction, attributed to the amorphous SiN substrate or amorphisation of the perovskite crystal. Weak diffraction reflections at $k$ = 0.26, 0.24 and 0.26 Å$^{-1}$ are also visible. These reflections are indexable to the 011, 010 and (01$\bar{1}$) peaks of PbI$_2$ at the [$\bar{1}$00] zone axis within 0.01 Å$^{-1}$ error, as shown in Figure 2d. PbI$_2$ can form by stacking PbI$_2$ layers in many possible ways, resulting in different polytypes.[41] We note the reflections observed here suggest the formation of the 4H polytype, in contrast to other polytypes such as 2H or 6H reported in other studies.[20,23] Despite some ambiguity in classifying the exact $PbI_2$ polytype in literature, these phases are remarkably similar and the formation of small precipitates of $PbI_2$ is unambiguously observed (see indexation in Figure S5).[41]

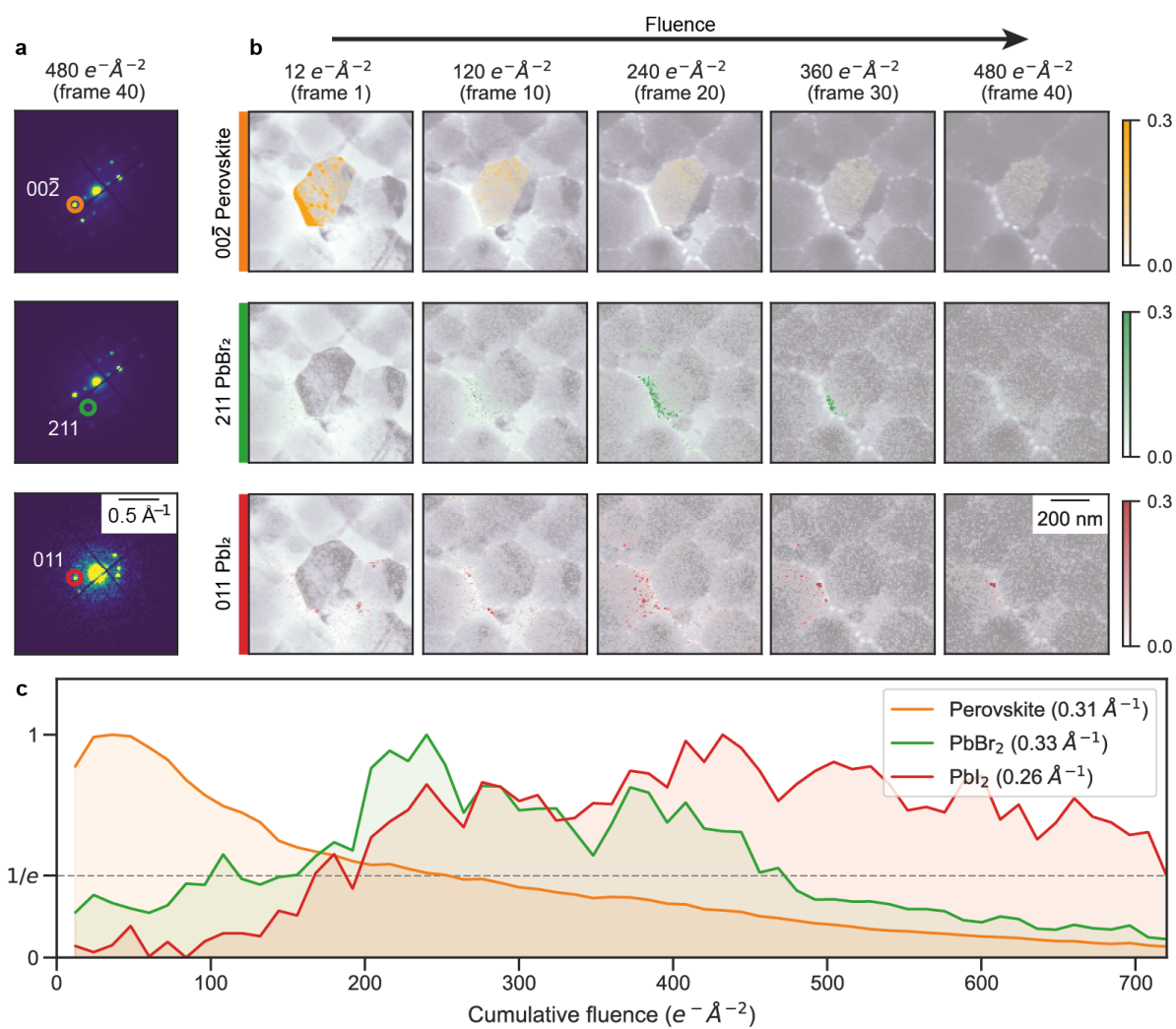

**Figure 3: Mapping the evolution of the degradation species in a $(FA_{0.79}MA_{0.16}Cs_{0.05})Pb(I_{0.83}Br_{0.17})_3$ film.** a) The diffraction patterns at the 40$^{th}$ frame, taken from region 1 in Figure 2 for the perovskite and PbBr$_2$ series, and from region 2 in Figure 2 for PbI$_2$. For each diffraction pattern, a dark-field virtual aperture is selected. b) vDF images of the respective virtual apertures placed at: the $(00\bar{2})$ perovskite reflection (orange), the (221) PbBr$_2$ reflection observed within the grains (green), and the (011) PbI$_2$ reflection localized at the high-angle grain boundary (red). The color scale range is set to 30% to better show the small changes in intensity. All vDF images are superimposed on top of the respective vBF images. c) Intensity profiles taken from the perovskite $(00\bar{2})$, the PbBr$_2$ (221), and the PbI$_2$ (011) reflections. The 1/e intensity threshold is shown with a dashed line.

To further understand the evolution of phases in Figure 2, we use SED to spatially map the crystallographic changes. **Figure 3** shows the evolution of the vDF images created for the perovskite $(00\bar{2})$ reflection, the $(211)$ $PbBr_2$ reflection, and the $(011)$ $PbI_2$ reflection (Figure 3a). While the vDF image intensity for the perovskite diffraction peak fades upon exposure, the vDF images of the additional reflections can spatially pinpoint the origin of these changes (Figure 3b). The lead halides appear to be spatially anticorrelated: epitaxial 4H-$PbI_2$ precipitates grow at the grain boundary and the $PbBr_2$ reflections appear at the opposite side of the boundary within the perovskite grain. At extreme cumulative fluences of >480 e$^-$ Å$^{-2}$ (40 frames), most diffraction from crystalline species has faded, so the vDF images are dominated by diffuse scattering and noise. Some additional vDF images of $PbI_2$ precipitates forming at the grain boundaries are identified in Figure S11 and S12, though we note that any $PbI_2$ trace crystallites that are too small may be undetectable in SED. The $PbI_2$ features are extremely local, only detectable across a few scanned pixels (~5-30 nm).

These changes are local to the bottom-left corner of the grain, suggesting that lead halides nucleate non-uniformly across the grain. Degradation does not appear to originate from the twin boundaries, consistent with theoretical predictions, whereby the octahedron face-sharing present at the twin boundary stabilizes the twins.[42,43] Therefore, not all defects affect degradation in the same way.

Intensity profiles for the perovskite $(00\bar{2})$, the $(211)$ $PbBr_2$, and the $(011)$ $PbI_2$ reflections are plotted in Figure 3c. These profiles suggest that the critical fluence for this mixed cation composition, the fluence at which the original diffraction intensity has reached ~1/e relative to its initial value, is ~200 e$^-$ Å$^{-2}$. At this critical fluence, the lead halide phases start to emerge. The critical fluence reported here is in close agreement with those by Rothmann *et al.* for the pure $FAPbI_3$ composition.[23] We investigated other grains from adjacent regions, all of which exhibited similar progressive grain amorphisation with an equivalent critical fluence (see Figure S13).

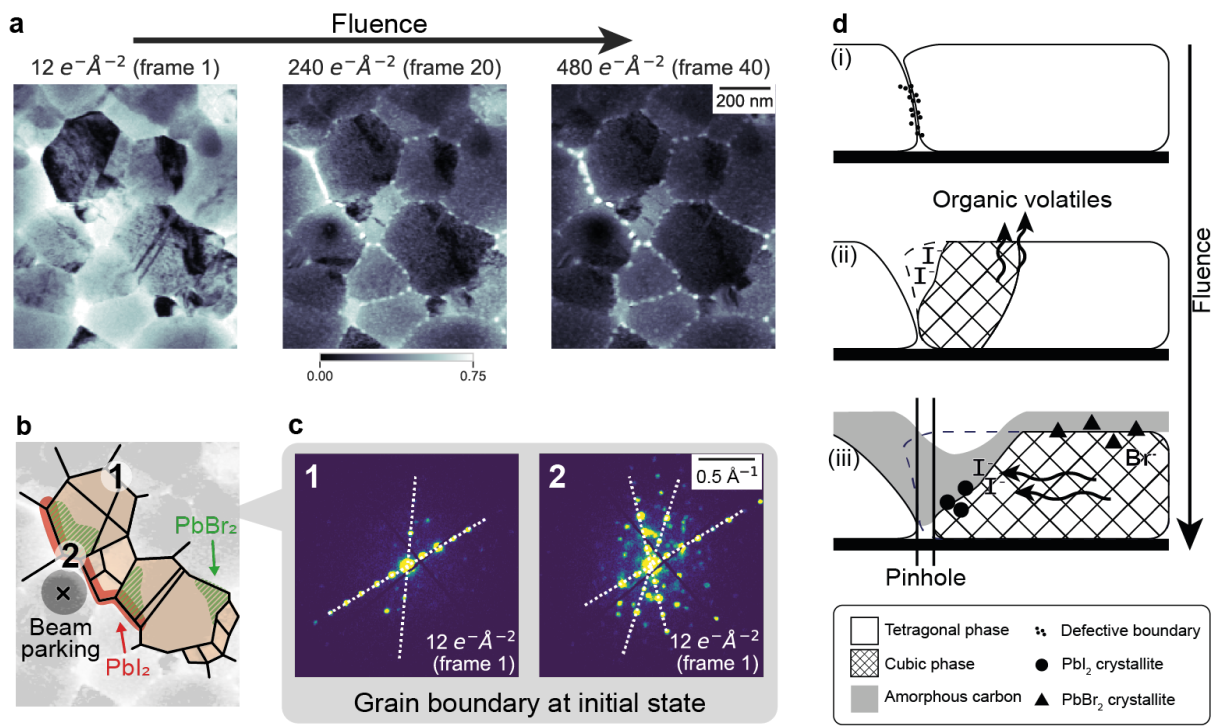

**Figure 4: The grain boundaries in a (FA$_{0.79}$MA$_{0.16}$Cs$_{0.05}$)Pb(I$_{0.83}$Br$_{0.17}$)$_3$ film affect the degradation pathway.** a) Evolution of vBF images of various grains and its surrounding grains after up to 480 e$^-$ Å$^{-2}$ fluence, showing stark changes in the morphology. b) Schematic of multiple perovskite grains from which PbI$_2$ and PbBr$_2$ species are observed. The regions from which diffraction patterns are extracted are labeled as 1 and 2. c) Diffraction patterns for two high-angle grain boundaries: in region 1 a grain boundary exhibiting the reflections only attributed to the perovskite phase of the grains at either side of the boundary, and in region 2 a grain boundary also exhibiting additional reflections and diffuse scattering, typical of amorphous defective phases. d) Schematic of the degradation mechanisms and factors identified during degradation at high fluences, in cross-section view.

The nature of the grain boundaries affect the degradation pathway. High-angle grain boundaries, labeled as 1 and 2 in **Figure 4**, are associated with the nucleation and growth of pinholes. Not all grain boundaries exhibit pinhole formation to the same extent, exemplified by region 2 being more altered than region 1 (Figure 4a). The initial diffraction pattern at this region reveals region 2 to show not only reflections attributed to the perovskite adjacent grains, but also additional reflections and diffuse scattering. Although these additional reflections, already visible in the first frame, are difficult to index, they suggest the presence of defects.[44] These defective high-angle boundaries can trigger faster degradation and larger morphological variations than less defective boundaries (region 1).

In the schematic in Figure 4e, we propose a degradation mechanism whereby iodine segregation towards the grain boundary leaves bromine rich areas within the perovskite grains, which form $PbBr_2$ at the surface to balance the iodine deficiency. Both $PbI_2$ and $PbBr_2$ are more thermodynamically stable and have higher thermal conductivity than the parent hybrid perovskite. The transformation dissipates energy and heat more effectively. These processes happen in conjunction with the loss of the more volatile $I_2$ and organic moieties. These changes result in a tetragonal to cubic phase transition, the formation of small $PbI_2$ crystallites at the grain boundaries with pinholes, and the redeposition of some organics as a thin amorphous organic film on the surface of the specimen.

A grain near the [001] zone axis, which exhibits superlattice reflections attributed to the $BX_6$ corner-sharing octahedra tilted away from perfect cubic symmetry towards the tetragonal perovskite phase,[38] shows a progressive loss of these superstructure reflections (see Figure S14). These changes upon electron exposure suggest a gradual transformation from tetragonal to cubic structure, expected with the loss of the organic cations and in agreement with the proposed degradation mechanism.

In this SED experiment the beam parking position was set at the center of the image frame, as shown in Figure 3b, causing overexposure within the field of view. This may cause additional damage to the grain under beam parking and its boundaries, and trigger faster degradation. However, similar degradation pathways are observed in grains that are not adjacent to the beam parking position. For example, we see additional $PbBr_2$ reflections emerging after a fluence of ~252 e$^-$ Å$^{-2}$ from another grain oriented near [100] (Figure S11 and S12).

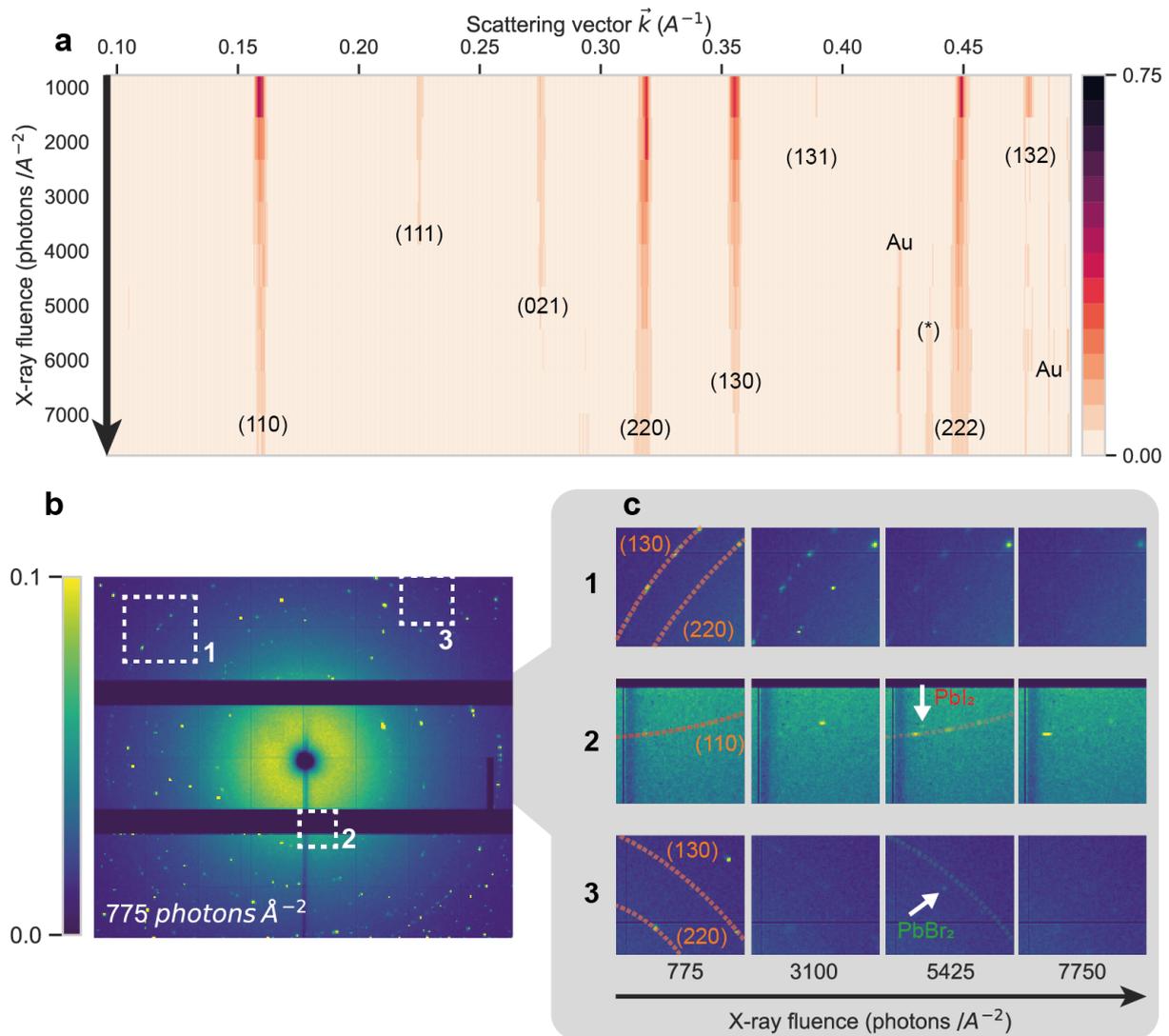

**Figure 5: Probing the evolution of the local structure of a region in a $(FA_{0.79}MA_{0.15}Cs_{0.06})Pb(I_{0.85}Br_{0.15})_3$ halide perovskite film under X-ray exposure using nXRD.** a) Radially integrated nXRD evolution of a perovskite film over accumulated illumination over an area of 4 μm². Peaks are indexed to the P4/mbm perovskite phase. Peak marked as (*) cannot be indexed to perovskite and corresponds mainly to the degradation phases. To analyze the weak reflections lost during radial integration, 2D nXRD diffraction patterns are shown in b), with some areas of interest denoted by numerals. c) The zoomed-in evolution of nXRD in 2D elucidates crystal tilting in region 1, the weak appearance of (002) 4H-PbI₂ in region 2, and the appearance of (211) PbBr₂ reflections in region 3. These findings are similar to the changes observed in SED.

---

Finally, we perform analogous scanning probe synchrotron nXRD measurements to acquire a similar 5D dataset by combining a time series stack of multiple nXRD frames. Such acquisition allows us to study the interaction of highly converged 20 keV X-rays on nominally equivalent triple cation composition perovskite films as previously presented for electrons. The use of nXRD allows for close comparison with SED because it similarly involves scanning over a localized region of interest. Unlike SED, this technique cannot access subgrain features under these conditions, as the X-ray beam diameter is ~150 nm. **Figure 5a** shows the spatially averaged 1D diffraction patterns across a scanned area of ~4 µm$^2$ for X-ray photon fluences up to 7,750 photons Å$^{-2}$.

The main diffraction peaks are indexed to the same tetragonal P4/mbm perovskite phase as in SED. Some crystallographic plane reflections are more prevalent and decay more slowly, which can be linked to the crystal structure. (110), (220), (130) and (222) planes are the brightest reflections. The diffraction intensity at these planes is strongly affected by the loss of the halogen atoms from the unit cell since these planes cut mainly across halide positions (see the simulation in Figure S15). In contrast, (111), (021) and (131) planes diffract less strongly, and their relative intensity is more affected by the loss of the organic cations since all contain A+X ion sublattices. In fact, the (220) and (130) planes exhibit the slowest decay rate (see Figure S15), suggesting the rapid loss of the organic cation followed by the slower loss of the halides. The loss of I and Br can be further analyzed from the nXRF, acquired simultaneously to the nXRD data, in which the $I_{La}$ signal systematically drops before the $Br_{Ka}$ signal (see Figure S16), suggesting $I_2$ to be more volatile than $Br_2$.[45]

The (110), (220) and (130) plane reflections brighten after 2-4 frames (~1,400-2,800 photons Å$^{-2}$). A closer examination of these reflections in the 2D patterns (Figure 5c, region 1) reveals the appearance of additional Bragg reflections of the crystal planes, attributed to the crystals tilting closer to the zone axis, similar to the SED findings in Figure 1 albeit in nXRD we are sampling multiple grains.

Some weak and broad reflections appear after 5 frames (4,650 photons Å$^{-2}$) in the range $\sim 0.436 \pm 0.003$ Å$^{-1}$ (Figure 5a, marked with *). These reflections are not indexable to the perovskite structure and can be used to arbitrarily determine the X-ray critical exposure of hybrid perovskite systems. This critical exposure value is one order of magnitude higher than previously reported values for $MAPbBr_3$ single crystals ($10^2$ photons Å$^{-2}$), at which point the scattering image had changed by more than 10%, but it is lower than for $CsPbBr_3$ single crystals ($10^4$ photons Å$^{-2}$).[25] We also performed the same synchrotron diffraction degradation experiments using a static 17.2 keV X-ray box-beam with a wider illumination area of ~1 µm$^2$ (see Figure S17). The non-localized nature of the exposure required 2-3 orders of magnitude higher fluence (~5·10$^6$ photons Å$^{-2}$) to acquire similar diffraction information and produce a similar degradation pattern as for nXRD.

The reflection at ~0.436 ± 0.003 Å$^{-1}$ matches the (110) 4H-PbI$_2$ plane, but also matches phases like PbBr$_2$ or poly-phases. A closer examination of the 2D diffraction patterns reveals extremely weak additional reflections appearing, indexable to both (002) 4H-PbI$_2$ and (211) PbBr$_2$ (Figure 5c region 2 and 3, respectively). As these reflections are lost during radial integration, it is critical to analyze the full 2D data sets (see more examples in Figure S18).

Our combined results offer an understanding of the changes induced by the interaction of electrons and X-rays with soft perovskite semiconductors. Each diffraction technique used offers some advantages over the other. nXRD provides excellent resolution in reciprocal space, but lower resolution in real space. X-ray based techniques are extremely useful to discern crystal structures with similar features. Since many of the degradation changes occur on the nanoscale, nXRD in general is unable to spatially resolve such changes.

In contrast, SED offers excellent spatial resolution and allows mapping of crystallographic changes at the nanometer scale. However, in SED, points in reciprocal space are broadened by a shape factor and spread out forming disks, due to the use of thin samples and a finite convergence semi-angle, respectively. Multiple of these diffraction discs are intersected by the lower curvature of the Ewald sphere, compared to X-rays, limiting the resolution in reciprocal space of SED. Due to the similarity between pristine and intermediate perovskite degradation phases such as the 2H, 4H and 6H polytypes,[28] SED is not able to unambiguously assign similar structures. Moreover, due to the constrained acquisition geometry, not all grains produce interpretable SED patterns near a low-order zone axis. Further studies using precession electron diffraction, which offers more complete integration of the diffraction space, could overcome the latter constraint if performed under low fluences.[46]

The different nature of electrons and photons result in distinctive interactions with the specimen. For example, when electrons propagate through a material, they undergo elastic Bragg scattering and inelastic scattering. For a 200 kV beam, each inelastically scattered electron transfers ~27 eV per scattering event to the specimen (see Table S1). This results in knock-on damage, radiolytic damage, and local heating.[13] For a 200-nm thick film of the perovskite composition studied here at 200 keV, we estimate the probability of inelastic scattering to be twice that of no inelastic scattering (calculated with the Poisson model in the SI).[47] The ratio of elastic versus inelastic cross-sections is inversely proportional to atomic number (proportional to $19 \cdot Z^{-1}$ for electrons of hundreds of keV).[47] For the triple cation perovskite composition, similar elastic and inelastic scattering probabilities are expected, based on the effective atomic number (Z= 36.4). However, carbon atoms will contribute more to, and be more affected by, inelastic scattering than iodine or bromine, by a factor of 8 and 6, respectively. This is consistent with the damage mechanism proposed, in which the organic cations are the weaker species, and the halides restructure chemically to more stable intermediates. The effect of the inelastic cross-sections can be reduced by using thinner specimens or higher acceleration voltages, both of which will reduce the accessible diffraction information. A fine balance between these parameters is key. For the X-ray photons of tens of keV, the perovskite film is virtually transparent (see calculations in the SI). However, inelastic scattering is also always present in XRD, with a ratio of inelastic to elastic scattering cross-sections also inversely proportional to atomic number by $50 \cdot Z^{-1.7}$ for 20 keV photons. These ratios highlight the higher need to account for photon beam damage when X-rays are used on hybrid materials, especially for prolonged exposures (see estimations in the SI and Figure S19).

Similar mechanisms of interaction and damage can be seen with both electron and X-ray probes. Grains tend to tilt by a few degrees towards the zone axis after the first frames of acquisition (~100 e$^-$ Å$^{-2}$ and ~1,400-2,800 photons Å$^{-2}$). We attribute these changes to microstructural changes on the nanoscale, yet beam-assisted annealing of the pre-strained films on the SiN support membrane cannot be excluded. These findings are important as they place constraints on the use of such techniques for elucidating the pristine microstructure of halide perovskites. Radiation exposure also leads to loss of crystallinity in the grains and degradation of the perovskite structure, with a critical fluence of <200 e$^-$ Å$^{-2}$ and <5,000 photons Å$^{-2}$ for SED and nXRD, respectively. The degradation mechanism proposed in this work is in agreement with the HR-STEM work on FAPbI$_3$ evaporated films by Rothmann *et al.*, in which PbI$_2$ precipitates epitaxially grow from the perovskite.[23] However, we observe crucial differences given by the more complex perovskite composition used in this study, in which simultaneous formation of PbBr$_2$ is also observed at the grains, likely aided by higher I volatility compared to Br. These findings can be extended to the radiation damage from a high-energy X-ray nanobeam, which also results in the emergence of lead halide species upon loss of the organic moieties across the whole scanned area (Figure S16). In fact, the presence of the organic cation species at the A position has been proposed to trigger degradation processes towards PbI$_2$ as opposed to the Pb$^0$, mostly only seen from inorganic compositions, as the organics can undergo radiolysis more easily than the $Pb^{2+}$ cation upon electron and X-ray illumination.[22,48]

The degradation mechanism proposed here contrasts with the degradation pathways observed from light-soaking. Photocarriers produced during light soaking can trigger the formation of $I_2$ via redox-mediated reactions, leaving metallic $Pb^0$ and pinholes behind. Such changes are proposed to be seeded from crystallographic and compositional impurities.[44] Light soaking can also trigger Cs segregation,[49] which is not observed in the nXRF mapping during high-energy X-ray exposure in this report (see Figure S16). The use of high-energy beams results in the study of faster degradation pathways of perovskites from its pristine state, rather than the study of early-stage degradation from impurities caused by the distinctively lower-fluence visible light. We note that large charge carrier densities are estimated to be created during acquisition, ranging around $10^{16}$ to $10^{19}$ $e^-$-$h^+$-pairs $cm^{-3}$ for SED, and $10^{19}$ $e^-$-$h^+$-pairs $cm^{-3}$ for nXRD (see estimations in the SI). Such large localized hot carriers can trigger radicals, multiparticle recombination and thermalisation processes on top of the knock-on damage, radiolytic damage, and local heating caused by the high-energy particle beam. Therefore, such high carrier concentrations can promote degradation of the perovskite in ways distinct from those encountered under illumination at 1 sun ($10^{14}$ to $10^{16}$ carriers $cm^{-3}$, mostly generated at band edge), likely aiding in the radiolysis of the organic cations in the perovskite.[48]

The radiation hardness described here suggests good resilience of these materials to degradation from electron radiation in space. Assuming radiation damage of hybrid perovskite being dependent on the total dose rather than on the dose rate,[11,12] the critical fluence for electrons would be accumulated after around ~2,000 years in the Earth orbit or ~200 years at harsher orbits like Jupiter (see Figure S20). We note the differences between electron radiation in microscopes and space, such as the energy spectrum of space radiation extending up to 1 MeV energies, and the generally lower radiation fluxes in orbit.[6,9] Combined with the reported excellent proton radiation hardness,[6,8] this study once more places halide perovskites as promising candidates for space photovoltaic applications. Further studies should also include the effect of cryogenic temperatures on the degradation of halide perovskites.

## 3. Conclusion

Our conclusions have implications for understanding the changes on the nanoscale of halide perovskites upon interaction with high-energy radiation, in particular electrons and X-rays. We have identified local changes on the nanostructure at extremely low radiation exposure, leading to changes in the orientation of the grains. At high radiation exposure, degradation of the perovskite phase involves iodine migration, producing decomposition to lead halide species. $PbBr_2$ reflections appear within perovskite grains that exhibit $PbI_2$ crystallites growing at their grain boundaries. Specifically, regions showing high-angle defective grain boundaries are important for degradation. Passivation and growth strategies targeting the removal of such high-angle defective grain boundaries will be critical for the further mitigation of halide perovskite instabilities, especially for applications vulnerable to high-energy radiation. Finally, the findings reported here provide a further understanding of the exposure limits of high-energy electron and photon beams, being <200 e- Å-2 at 200 keV and <5,000 photons Å-2 at 20 keV, respectively. These critical radiation exposures are crucial for any characterisation technique studying the nanoscale of halide perovskites and, more generally soft semiconductors, as well as to further understand the challenges that perovskite solar cells still face for space applications.

## 4. Methods

*Sample preparation:* All procedures were followed inside an inert $N_2$-gas glovebox. The perovskite precursors solutions were prepared by first dissolving FAI (1.0 M) and MABr (0.2 M), $PbI_2$ (1.1 M), and $PbBr_2$ (0.22 M) in a mixture of anhydrous DMF and DMSO (4:1 v:v). A solution of CsI (1.5 M in DMSO) was then added to the precursor solution as 5% of the total volume, yielding a $(FA_{0.79}MA_{0.16}Cs_{0.05})Pb(I_{0.83}Br_{0.17})_3$ perovskite precursor. Lead halide precursors were supplied by TCI, organic compounds were supplied by Greatcell Solar, CsI and solvents were supplied by Sigma.

To fabricate thin electron-transparent specimens for SED, the perovskite solution was diluted in anhydrous DMF: DMSO 4:1 (v:v) in a 2:1 ratio of diluent to precursor solution, and spin-coated on SiN TEM grids with a 30-nm thick low-stress amorphous $Si_3N_4$ membrane window (NT025X, Norcada). Spin-coating was followed in two steps at 2,000 and 6,000 rpm for 10 s and 35 s respectively, with 20 μl of chlorobenzene added 30 s before the end of the second step. The films were then annealed at 100 °C for 1 h, yielding films of ~200 nm.

To fabricate thin films for the nXRD studies, the same protocol was used to prepare an almost identical perovskite precursor solution of $(FA_{0.79}MA_{0.15}Cs_{0.06})Pb(I_{0.85}Br_{0.15})_3$. The precursor was then spin-coated on X-ray transparent windows (Norcada, NX7100C), following the same protocol as for the SED samples, and then annealed, yielding films of ~500-600 nm.

*SED:* During SED microscopy, a 2D electron diffraction pattern is measured at every probe position of an electron beam in STEM mode. SED data was acquired on the JEOL ARM300CF E02 instrument at ePSIC (Diamond Light Source, Didcot-Oxford, UK). A monolithic Merlin/Medipix direct electron detector with 4 back-contacts was used to acquire fast low-dose SED. These direct electron detectors allow for better SNR under lower doses due to the superior quantum efficiency compared to traditional CCD. The detector was set to 6-bit, to maintain the targeted electron fluence and fast acquisition readout rate. In the initial frames of the acquisition, some diffraction patterns exhibited saturation at the direct beam position, due to the bit-depth limitations of the direct electron detector imposed by the fast low-dose acquisition conditions. Such saturated pixels are occasional but visible as the brightest pixels in Figure 1b. The beam blanking after each frame was performed manually, always at the same controlled localized position and after a reaction time of ~500 ms. An acceleration voltage of 200 keV, nanobeam alignment (convergence angle ~1 mrad), electron probe ~5 nm, probe current ~3.59 pA, scan dwell time 1 ms, and camera length 20 cm. Post-processing of SED diffraction data was done using pyXem 0.12 (an open-source Python library based on HyperSpy-based for crystallographic electron microscopy).[50]

We refer to virtual bright-field images (vBF) as the images reconstructed from taking the intensity integrated solely from the direct beam as a function of probe position, thus only containing information from the electrons recorded at zero scattering angle. Contrarily, virtual dark-field images (vDF) are reconstructed from only taking the intensity from a Bragg-diffracted spot from the 2D diffraction pattern as a function of probe position, thus containing information solely on electrons scattered to specific Bragg angles. All virtual apertures for vBF mapping were set to $4 \cdot 10^{-3}$ Å$^{-1}$ (2.4 mrad of semi-angle of detector), and for vDF were set to $2-4 \cdot 10^{-3}$ Å$^{-1}$ at diffraction semi-angles ranging between 0 and 70 mrad.

All diffraction patterns were distortion corrected and calibrated with an Au cross grating. The drift in the beam position of the non-scattered beam was corrected and centered for all frames using cross-correlation with a subpixel factor of 10. To display diffraction planes that match real-space features, all diffraction patterns shown in the main text were rotation corrected using a $MoO_3$ calibration sample. Dead pixels and detector junctions were masked. No sample drift correction was necessary. Finally, the background noise in all diffraction patterns was filtered, setting a threshold of detector pixels with single counts to zero.

*nXRD:* Scanning nano-XRD data was acquired at the synchrotron beamline I14 of the Diamond Light Source (Didcot-Oxford, UK). 2D diffraction patterns were recorded at each stage position as the specimen was moved through the X-ray beam. An array of 3 Medipix 2048 x 512-pixel arrays in transmission mode (Excalibur 3M) was used to acquire diffraction data. To examine the effect of repeated maps on the same region of a film, a scan loop was measured within a 5 x 5 μm region until most of the diffraction spots were no longer visible. A 20 keV monochromatic X-ray beam ($\lambda$ = 0.619 nm) was used, focused to ~150 nm. Samples were mounted onto the I14 standard sample holder and measured under a local flow of dry $N_2$ to suppress perovskite degradation that is accelerated by moisture and oxygen. 2D sum patterns across the scanned region were used without processing. 1D patterns were radially integrated using the Data Analysis WorkbeNch (DAWN) with a $CeO_2$ calibration standard and a mask to remove dead pixels and detector edges.[51] The 1D sum pattern for each frame was cropped from 0.6 Å$^{-1}$ onwards and Origin Pro was used to subtract a baseline found from 20 baseline anchor points found via the 2$^{nd}$ derivatives and connected by line interpolation. Simultaneous X-ray fluorescence (nXRF) data was acquired with a 4 element Si drift detector. The nXRF maps were aligned using the $Au_{La}$ peaks and the summed signals across the scanned area were found for $Pb_{La}$, $I_{La}$, and $Br_{Ka}$.

All diffraction patterns in this work are reported in 'ordinary' wavevectors ($k = 1/d$). Note that these units differ from the "angular" wavevectors definition ($q = 2\pi/d$) by a factor of $2\pi$, which is often used in X-ray crystallography literature.

*Crystal structure library:* All diffraction patterns from this work can be closely indexed to a pseudo-cubic/tetragonal perovskite unit cell (P4/mbm) with lattice parameters of $a = b =$ 9.00 and $c = 6.36$ Å. Despite the stoichiometry of the sample being the complex $(FA_{0.79}MA_{0.16}Cs_{0.05})Pb(I_{0.83}Br_{0.17})_3$ perovskite composition, it can be approximated as a simpler unit cell made from the predominant ion at each site (C, Pb and I) and with manually scaled lattice parameters to fit the experimental observations. Any variations that would occur in diffraction patterns between the simpler model used here and the more complex structure are minor and hard to discern in the reciprocal-space resolution of our SED technique (under acquisition conditions one pixel in the detector is 0.0046 Å$^{-1}$).

A series of different plausible degradation crystal structures were also considered, to attempt the indexation of some of the additional reflections appearing after radiation exposure. In general, the hexagonal 4H-PbI$_2$ crystal phase (P63mc, lattice parameters of $a = b = 4.56$ and $c = 13.96$ Å, COD ID: 9009140) and the orthorhombic PbBr$_2$ crystal phase (Pnam, COD ID: 1530324, lattice parameters of $a = 80.6$, $b = 6.54$ and $c = 4.73$ Å) were found to match the diffraction patterns discussed in this work. The structure files were retrieved from the crystallography open database.[52,53] However, other plausible degradation phases were also considered but did not match experimental data, such as perovskite intermediate polytypes, which are more extensively described in the SI, or the other 2H- and 6H-PbI$_2$ polytypes reported by others as a degradation product.[20,23] However, despite some ambiguity on the specific PbI$_2$ polytype across literature, since these phases are remarkably similar,[41] the formation of small precipitates of PbI$_2$ is observed.

All diffraction simulations were performed using Single Crystal 4 (CrystalMaker Software Limited), adjusting the simulation parameters to resemble the experimental data (200 keV, detector spot size 0.025 Å$^{-1}$, saturation 10, gamma 2). Simulation files in '.scdx' format can be found in the Supporting Data.

**Supporting Information**

Supporting Information is available from the Wiley Online Library or from the author.

**Acknowledgements**

J.F.O and C.D. acknowledge funding from the Engineering and Physical Sciences Research Council (EPSRC) Nano Doctoral Training Centre (EP/L015978/1). T.A.S.D. acknowledges funding from a National University of Ireland Travelling Studentship. J.F.O., T.A.S.D., and S.D.S. acknowledge the European Research Council (ERC) under the European Union's Horizon 2020 research and innovation programme (HYPERION, grant agreement no. 756962). S.D.S. acknowledges support from the Royal Society and Tata Group (UF150033). The authors acknowledge funding from the EPSRC (EP/R023980/1), and from the EPSRC Centre for Advanced Materials for Integrated Energy Systems (CAM-IES, EP/P007767/1).

SED studies were supported by the access to e02 at ePSIC (MG25250) and nXRD studies were supported by access to the I14 beamline (SP-20420). Mohsen Danaie and Julia Parker are acknowledged for their support during the acquisition and calibration of the SED and nXRD data, respectively.

**Author contributions:**

J.F.O. analyzed and interpreted SED, nXRD, and sXRD data. T.A.S.D. prepared samples and acquired the microscopy datasets and assisted with the interpretation of the data. D.N.J. and S.M.C. aided in the acquisition and interpretation of SED data. P.A.M. supervised D.N.J. and aided in the interpretation of SED data. H.S. assisted in the data acquisition of data. S.D.S supervised J.F.O. and T.A.S.D. and provided input on data interpretations. C.D. supervised J.F.O. and provided input on data interpretations. J.F.O. wrote the manuscript with input from all authors.

**Conflicting interests**

S.D.S. is a co-founder of Swift Solar, Inc.

# Supporting Information

**Unveiling the interaction mechanisms of electron and X-ray radiation with halide perovskite semiconductors using scanning nano-probe diffraction**

*Jordi Ferrer Orri,[1,2] Tiarnan A.S. Doherty,[1] Duncan Johnstone,[2] Sean M. Collins,[3] Hugh Simons,[4] Paul A. Midgley,[2] Cate Ducati,[2,*] and Samuel D. Stranks[1,5,*]*

[1]*Cavendish Laboratory, University of Cambridge, Cambridge, UK.*
[2]*Department of Materials Science and Metallurgy, University of Cambridge, Cambridge, UK.*
[3]*School of Chemical and Process Engineering & School of Chemistry, University of Leeds, Leeds, UK.*
[4]*Department of Physics, Technical University of Denmark, Denmark.*
[5]*Department of Chemical Engineering and Biotechnology, University of Cambridge, Cambridge, UK*
*Corresponding authors: cd251@cam.ac.uk, sds65@cam.ac.uk

# Contents

# Extended methods

**Static XRD sample fabrication:** To fabricate the samples for the static XRD studies, glass coverslips (18 mm x 18 mm, 0.13-0.17 mm thickness, Academy) were cleaned in acetone and isopropanol (10 minutes each) in an ultrasonic bath. The substrates were treated for 10 minutes in an oxygen plasma cleaner immediately before the spin-coating procedure and annealing thereafter. The same protocols as for the nXRD sample preparation were followed, yielding films of ~500-600 nm thickness.

**Static XRD:** Static synchrotron XRD data was acquired at the European Synchrotron Radiation Source (ESRF, Grenoble, France). A 17.2 keV monochromatic X-ray beam ($\lambda$ = 0.729 Å) was focused to a 'box beam' significantly larger than the individual grain size (~150nm), approximated as 1 $\mu m^2$, ensuring that there was a uniform flux intensity across the beam, and that a large number of grain orientations were sampled. The diffracted X-ray signal mas collected by a FReLON CCD camera placed 0.188 m from the sample. A sample was irradiated, forming a 2D diffraction pattern, and then an image was acquired every 0.5 s (with a dwell time between frames of 1.14 s) until the diffraction spots were no longer visible. The incoming flux was indirectly measured using an ion chamber with $N_2$ gas and with a picometer current detector.

From the 2D detector images, the Python Fast Azimuthal Integration tool (pyFAI) was used to azimuthally integrate to 1D XRD patterns.[1] The standard detector geometry distortion file from the ESRF was used to calibrate the radial integration, which was adjusted to match the tetragonal P4mbm perovskite phase (see Figure S9a). Finally, since the beam flux exhibited some fluctuations during the acquisition, each frame was rescaled accordingly to the normalised incoming flux and was background subtracted by fitting a linear background taken from the mean intensity values in the range before the first measured peak (0.10-0.15 Å$^{-1}$).

# Calibration of flux and fluence

**Note on definitions:** *Flux*: Number of particles per unit time (takes into account the beam source, such as e⁻ per second, photons per second…). *Fluence*: Number of particles per unit area, also known as flux density (takes into account the beam size as well, such as e⁻Å$^{-2}$, photons m$^{-2}$…). *Dose*: Amount of energy absorbed per unit mass (takes into account the sample-beam interaction, calculated in joules per kg or Grays (Gy)). Oftentimes, these definitions in the electron microscopy field are not referring to the standard particle Physics definitions. Fluence is often referred as "dose". This is because modelling the electron-beam-sample interaction is not trivial and requires many assumptions. Most electron microscopy literature states "dose" as e⁻Å$^{-2}$, as fluence.

This is in contrast to the X-ray microscopy field, which uses the standard definitions, and doses are always stated in Gray units. Note that sometimes administered dose is reported instead of the absorbed dose, value which does not take into account the sample-beam interaction.

Here, we report fluence for all techniques in units of particle (e⁻ or photon) per square ångström.

**SED:** The fluence for the SED experiment was calculated from the measured probe current during data acquisition. A probe current of 3.59 pA is equivalent to a flux of $2 \times 10^7$ e⁻ s$^{-1}$. All SED maps were acquired at a 1 ms dwell time using a Gaussian-shaped beam which can be approximated to a circular beam of ~5 nm diameter, slightly overestimating dose. These acquisition parameters result in a fluence of ~12 e⁻ Å$^{-2}$.

The simulation software CASINO v2.51, which simulates Monte Carlo electron trajectory in solids, was used to simulate the stochastic interaction volume for the electron beam.[2] This model is used because of the polycrystalline orientation nature of perovskite samples, and the total sample thickness. In such samples, the electron beam is likely to spread more significantly than in a well aligned single crystal, thus using the Monte Carlo method is more suitable than multi-slice or Bloch-wave simulation methods. Figure S1 shows the trajectory of $10^5$ electrons for a beam of radius 5 nm, for the default simulation parameters. The sample was modeled as 2 layers: a halide perovskite of 200 nm thickness and density 4.16g/cm$^3$, and a 30 nm Si$_3$N$_4$ layer. An interaction volume of ~90 nm radius contains the majority of the trajectories.

Conventional scan controls in STEM incorporate a left-to-right and row-by-row raster pattern requiring a 'flyback' time for stabilizing, prior to sequential row scanning. Moreover, beam blanking typically has a finite rise time. Given these issues, unavoidable in common STEM setups, we specifically parked the beam in a controlled position within the field of view, labelled with a cross in Fig. 1 in the main text. To understand the effect of beam parking, the excess irradiation fluence can be estimated. Beam blanking after each scan was done manually. Based on Human Reaction Time studies,[3] the average reaction time to click a button (task required to manually blank the beam at the end of a scan) is 273 ms, around 1/4th of a second. Assuming the reaction time during acquisition was no longer than 500 ms, the overexposure during acquisition of the beam parking position is 500 ms per scan. Given the 1 ms dwell time during scanning, the overexposure at the parking position is ~x500 larger. Such overexposure is localised within the circular beam of ~5 nm diameter, yet its effects extend around the ~90 nm radius interaction volume.

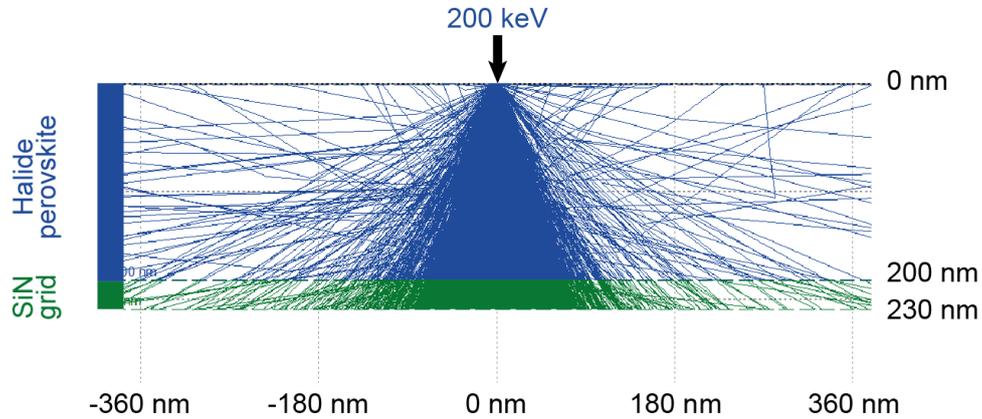

*Figure S1- Simulated electron beam trajectories using CASINO v2.51 at 200 keV for a layered sample containing a perovskite and SiN layers.*

**nXRD:** Flux is defined as photons per second. In the nXRD experiment, flux was indirectly measured using a calibrator diode (Si). The diode outputs a flux value in Amperes (A) per s, but has a responsibility factor. Taking as the diode responsibility conversion factor 0.108 ± 0.0005 A/W (from calibrations in the DIAMOND centre), the flux can be converted to photon power in Watts (W).

The rate of photons at a particular wavelength and power can be then calculated by using the equation for photon energy, $E = hc/\lambda$ (where h is Planck's constant and $\lambda$ = 0.619 nm from the beam energy of 20 keV), yielding a flux (photons per second).

However, the diode measurements will be slightly lower than the flux values on the sample, as the diode was placed within 200 mm away from the sample holder. X-ray attenuation by the air can be estimated using the mass attenuation coefficient of dry air for a 20 keV beam from the NIST database, as $\mu/\rho = 1.7 \times 10^{-2}$ cm$^2$/g.[4] Assuming a mass density of air of approximately 1.2 kg/m$^3$, an attenuation factor of $2 \times 10^{-5}$ cm$^{-1}$ is found. Using the Beer-Lambert law $I/I_0 = e^{-\mu x}$, we can estimate the X-ray attenuation across the 200 mm to be of 0.05%, therefore negligible.

The diode current measurement was $1.13 \times 10^{-7}$ A, which can be converted to a photon power of $1.1 \times 10^6$ W. Without the need to account for the flux attenuation by the air, a final flux value of $3 \times 10^9$ photons/s is calculated for each frame. Fluence (flux density) is defined as photons per unit area, and can be calculated from the X-ray beam size (which matches the rastered pixel size of the detector of 150 nm) and the dwell time per pixel (0.5 s), yielding a fluence of 775 photons Å$^{-2}$ for each frame.

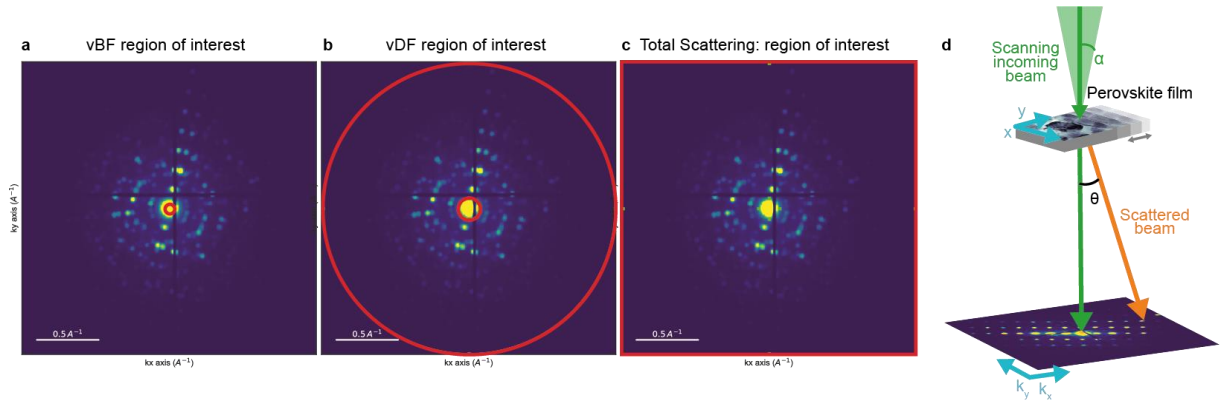

*Figure S2- Regions in the diffraction space from which a) vBF, b) vDF and c) total scattering images are taken to create the evolution movies attached in the SI. d) Schematic of the acquisition setup for the SED and nXRD system, in which the sample is moved through a static beam. For SED, the incoming beam is moving and the sample is static, whereas for nXRD the beam is static and the sample is moved. α is the semi-convergence angle and θ is the semi-angle of detection for the diffraction. For SED, the semi-convergence angle is 1 mrad. The range of the semi-angle of detection can be estimated: for a camera length of $2 \cdot 10^{-1}$ m and a pixel radius ranging from 0.04 (9 pixels in the detector) to 1.2 Å$^{-1}$ (256 pixels in the detector) for the vBF images in the main text and the vDF images in the SI, respectively. Given that the pixel size in the Merlin/Medipix detector is 55 μm, the θ ranges between 2 to 70 mrad (calculated using $tan^{-1}(\frac{|\vec{k}| \text{ in detector pixel length}}{CL: \text{ camera length}})$).*

# Electron diffraction simulation and indexation

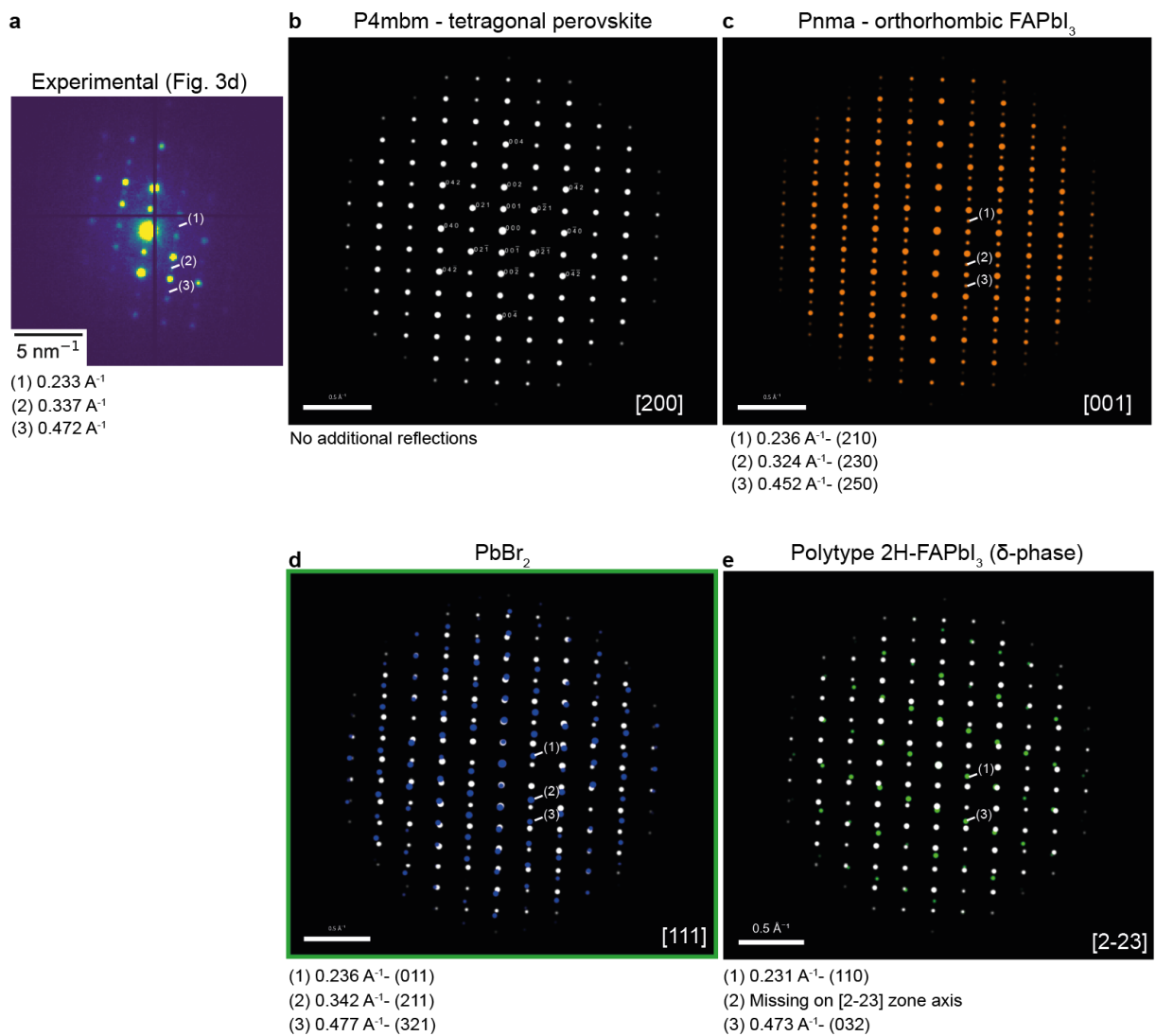

Figure S3- Indexation of the additional reflections seen in (a) the experimental data within the large tetragonal P4mbm perovskite-phase grains. Simulation of crystal diffraction patterns using SingleCrystal 4 for (b) the tetragonal perovskite phase, not showing any of the additional reflections, (b) orthorhombic FAPbI$_3$ phase, showing additional reflections which do not match the experimentally observed ones. These extra reflections at ½(hkl) originate from the tetragonal to the orthorhombic phase transition, which results in the tilting of the BX6 corner-sharing octahedral. (d) The simulation of PbBr$_2$ at the [111] zone axis closely matches the additional reflections in the experimental data set. (d) Any of the other plausible degradation phases, such as the FAPbI$_3$ delta phase shown as an example, did not match either.

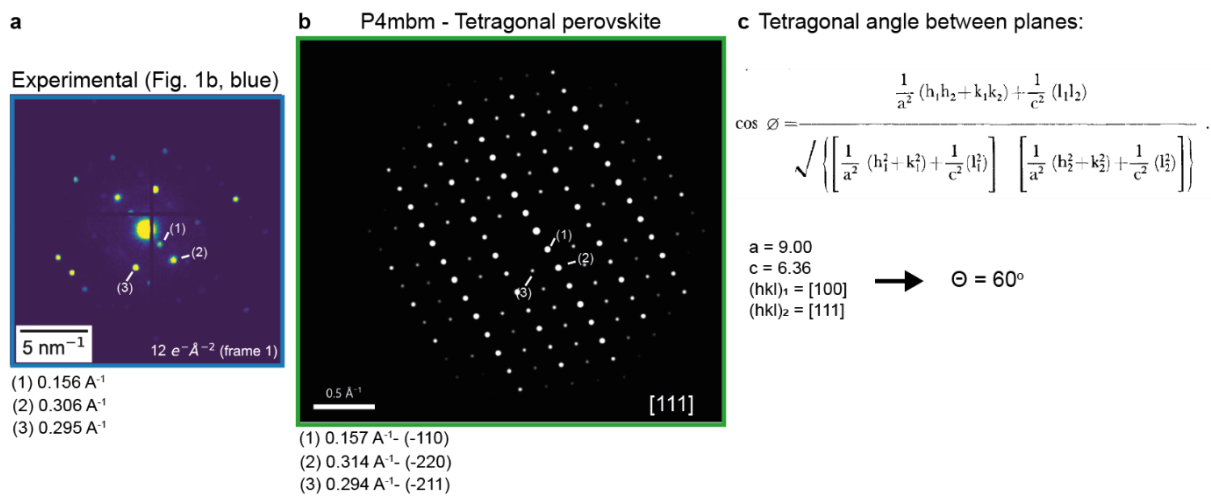

*Figure S4- Indexation of one of the adjacent grains from Fig 1b-c (in blue). (a) The experimental data, with some reflections marked as 1-3 which are indexable to the tetragonal perovskite phase at the [111] zone axis. (b) Simulation of crystal diffraction patterns using SingleCrystal 4, closely matching the experimentally observed reflections. (c) Estimation of the angle between the [100] and the [111] zone axes for a tetragonal unit cell of a=b=9.00 and c=6.36 Å$^{-1}$, estimating a high-angle grain boundary of ~ 60 degrees.*

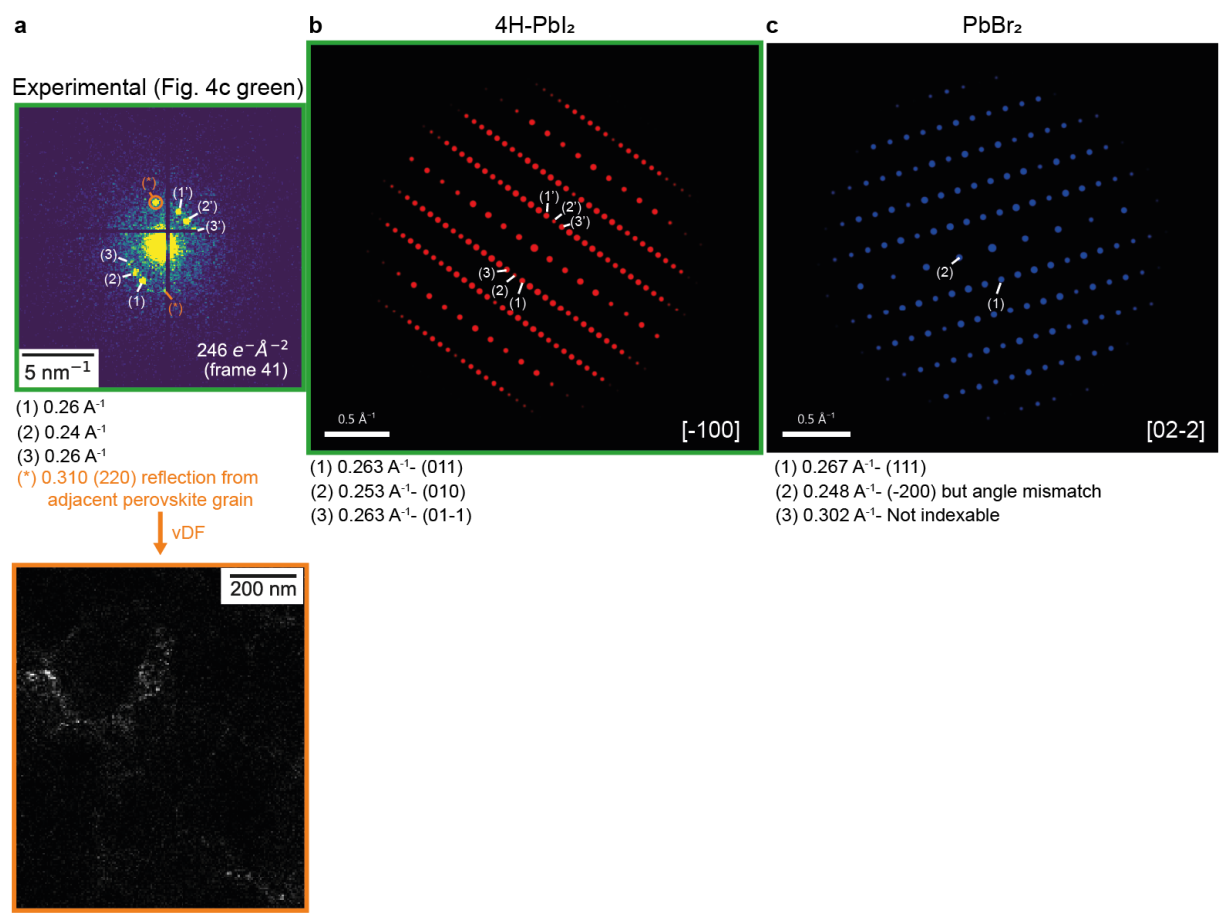

*Figure S5- Indexation of the additional reflections seen at (a) the experimental data at the grain boundaries, with some reflections marked as 1-3 which are not indexable to the tetragonal perovskite phase. Simulation of crystal diffraction patterns using SingleCrystal 4 for (b) the 4H-PbI$_2$ degradation phase, closely matching the experimentally observed reflections at [-100] zone axis. In contrast, other phases did not match such reflections, as illustrated by (c) for the PbBr$_2$ simulation, in which reflections do not match. The additional reflections marked as (*) in orange correspond the the pristine FAPbI$_3$ phase, as the vDF image shows.*

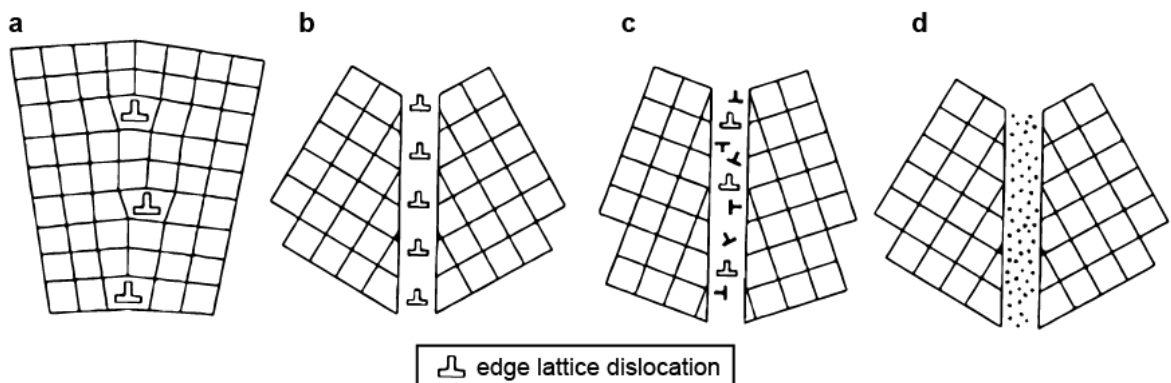

*Figure S6- Schematic of the types of grain boundaries. While atoms in the bulk phase form equilibrium crystal lattices at a minimum free energy of an infinite crystal, atoms at grain boundaries are arranged in a more disordered manner, imposed by arrangement restrictions from the adjacent crystal lattices. Different types of grain boundaries can be formed: (a) low-angle boundaries, (b) high-angle boundaries, (c) non-equilibrium grain boundaries containing high density of defects and (d) amorphous grain boundaries. Adapted from CRC Press.*[5]

# Grain tilting in-plane

The degree of the grain tilting in-plane can be estimated by fitting a Ewald circle passing though the brightest diffraction reflections and using Bragg's diffraction law:

$$\theta \sim \sin^{-1}\left(\frac{\lambda}{2 \cdot r^{-1}}\right) \text{ Equation S1}$$

Where r is the radius of the fitted circle (in the same units as the wavelength) and $\lambda = 2.5\ pm$ is the wavelength of the electron beam. For some grains of interest, were initially off zone axis are estimated in Figure S7. The angles at which the same grains are after 40 frames are also shown, mainly titled to zone axis [100] for the tetragonal phase. This titling is very slight, from 1-5 degrees, sufficient to allow these grains to relax to zone axis. The study of most of the indexable grains in the scanned region reveals that all grains show some degree of rotation if their initial state is not at zone axis. vDF images of these grain tilting are also shown in Figure S13.

Similar comparable results on other halide perovskite samples are observed. Figure S8 shows tilting observed in a SED crystallographic dataset of another triple cation double halide perovskite polycrystalline film composition grown on a SiN TEM grid. However, this SED experiment was acquired using a 300 keV JEOL ARM300F high-resolution STEM using a Merlin/Medipix direct detector with a single back contact (em19793-2 session, ePSIC, Diamond Light Source, UK). The observations in tilting of the original diffraction patterns upon initial radiation exposure are comparable to those reported in the main manuscript, making these observations representative across different films of same composition and across TEM acceleration voltages at 200-300 keV.

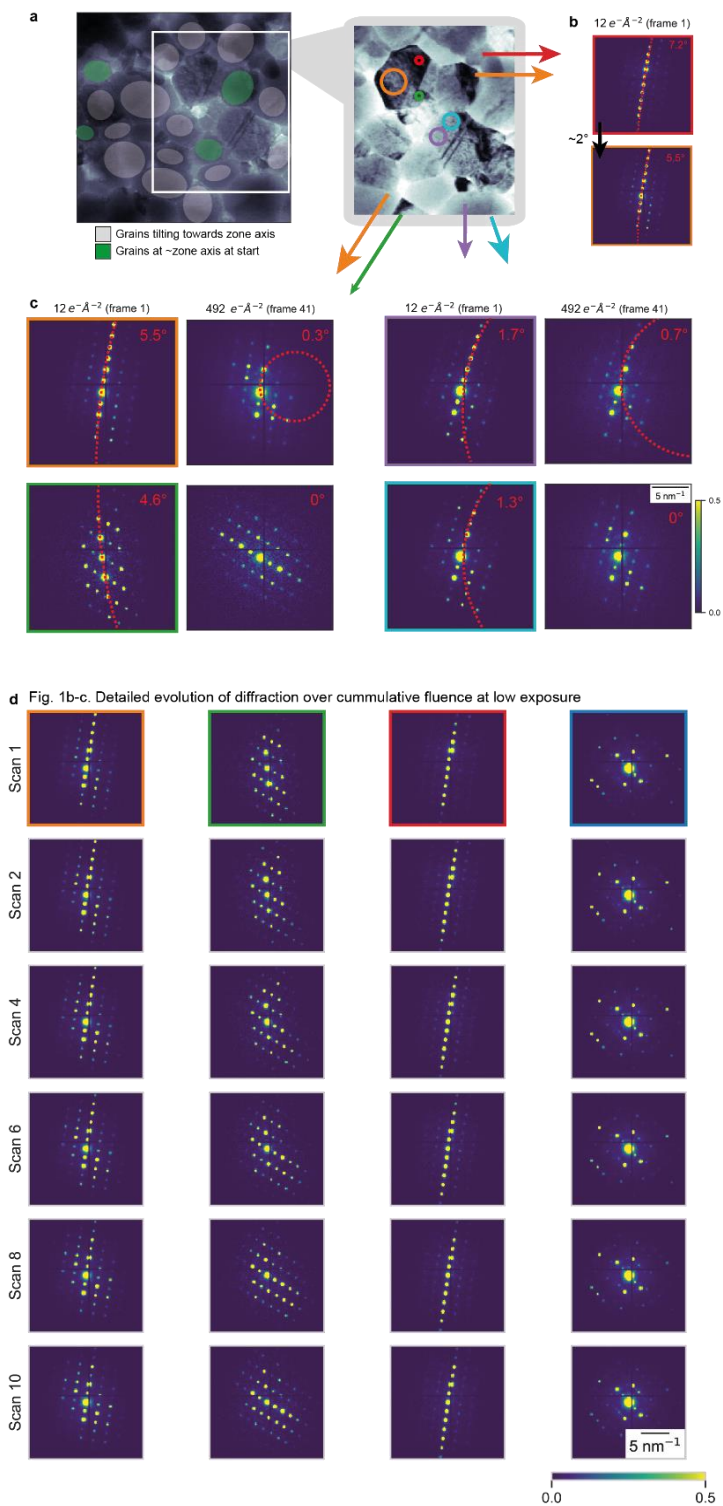

*Figure S7- a) In-plane grain tilting across the scanned region, most grains showing tilting towards the zone axis. b) Grain tilting quantification based on Bragg's equation for the orange and red regions (same grain) and c) the quantification of the tilt change over electron exposure. d) Extension to Fig. 1b and 1c in the main text, showing the evolution of the diffraction patterns at low cumulative fluence.*

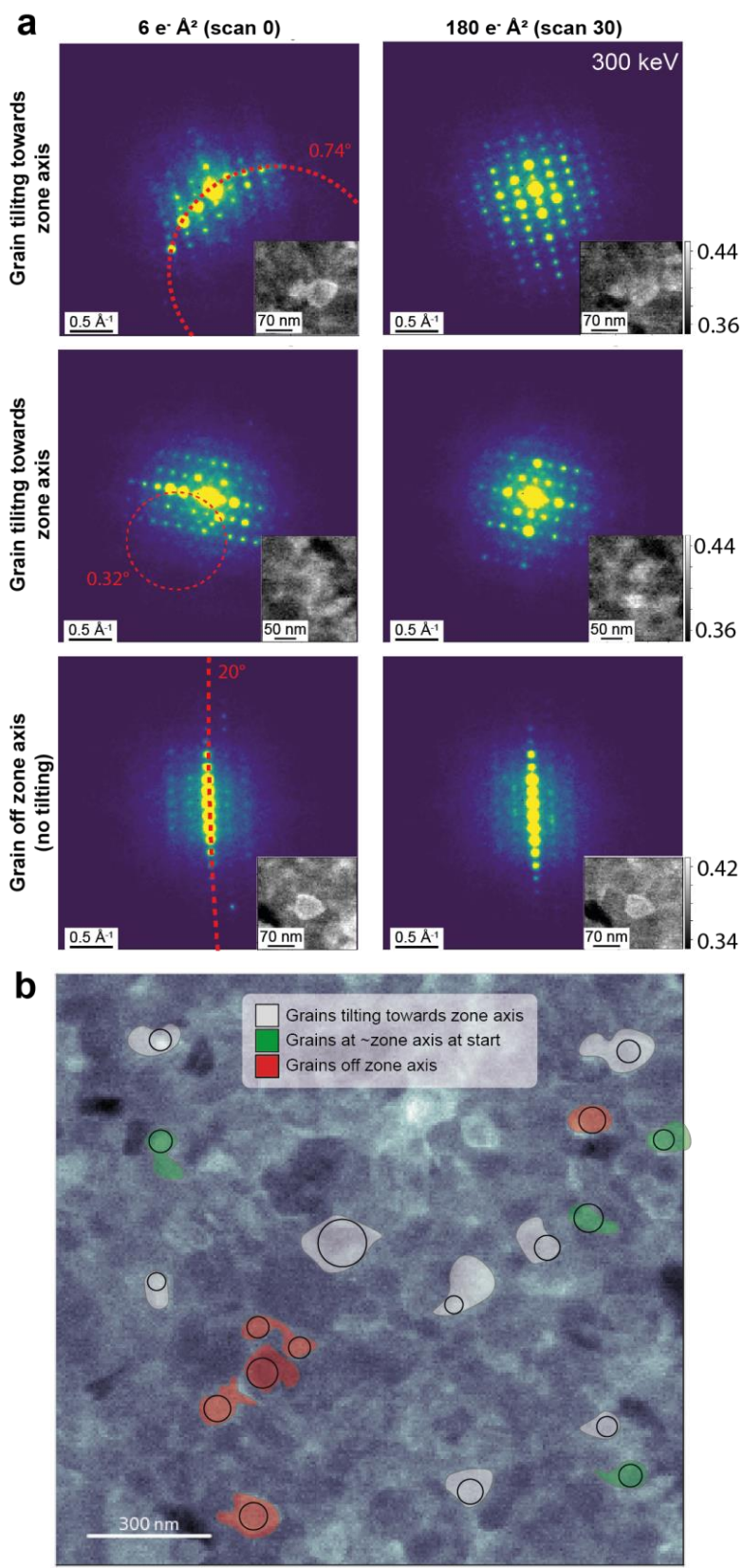

*Figure S8- a) In-plane grain tilting observed after little electron exposure, quantified based on Bragg's equation. b) Grain tiling across the whole scanned region, showing grains tilting towards the zone axis and other grains off zone axis. This SED crystallographic dataset was taken from a different halide perovskite film of same composition grown on a SiN TEM grid. It was acquired at 300 keV using a JEOL ARM300F high-resolution STEM with a Merlin/Medipix direct detector with a single back contact.*

# Degradation crystal structure library

Apart from the pristine tetragonal P4/mbm unit cell (Figure S9a), a series of different plausible degradation crystal structures were also considered in this study, to attempt the indexation of some of the additional reflections appearing after radiation exposure. Some of these phases are polytypes, or polymorphs which have identical close-packed planes but with different stacking sequences. Thus they can be classified by the Ramsdell notation, where the letter denotes the crystal system of the compound (C for cubic, H for hexagonal) and the number denotes the total number of layers contained in a unit cell. For the lead halides, the 4H-PbI$_2$ hexagonal (P6$_3$mc, lattice parameters of $a = b =$4.56 and $c =$13.96 Å, COD ID: 9009140) and the PbBr$_2$ orthorhombic (Pnam, COD ID: 1530324, lattice parameters of $a =$80.6, $b =$6.54 and $c =$4.73 Å) were used. These structure files were retrieved from the crystallography open database (COD) from the following references and are shown in Figure S9b.[6,7] Other degradation phases were also attempted, such as the intermediate polytype phases, reported by Gratia *et al.*,[8] which combine sequences of hexagonal and cubic closed-packed AX$_3$ stacks that result in a framework of face-sharing and corner-sharing BX$_6$ octahedra. They are called the 4H and 6H polytypes and are crystal structures that interface the photoactive α-phase (3R) with the yellow δ-phase (2H). However, the presence of these phases was inconclusive in this study.

Finally, to simulate the orthorhombic phase that would emerge from the additional degree of rotation in the BX$_6$ octahedra,[9] the tetragonal FAPbI$_3$ unit cell was taken as a model and tilted by a few degrees to obtain an orthorhombic FAPbI$_3$ (Pnma) unit cell with lattice parameters $a =$8.99, $b =$12.72 and $c =$8.99 Å as an approximation.

All diffraction simulation was performed using Single Crystal 4 (CrystalMaker Software Limited) adjusting the simulation parameters to resemble experimental data. Simulation files in ".scdx" format can be found in the Supporting Data.

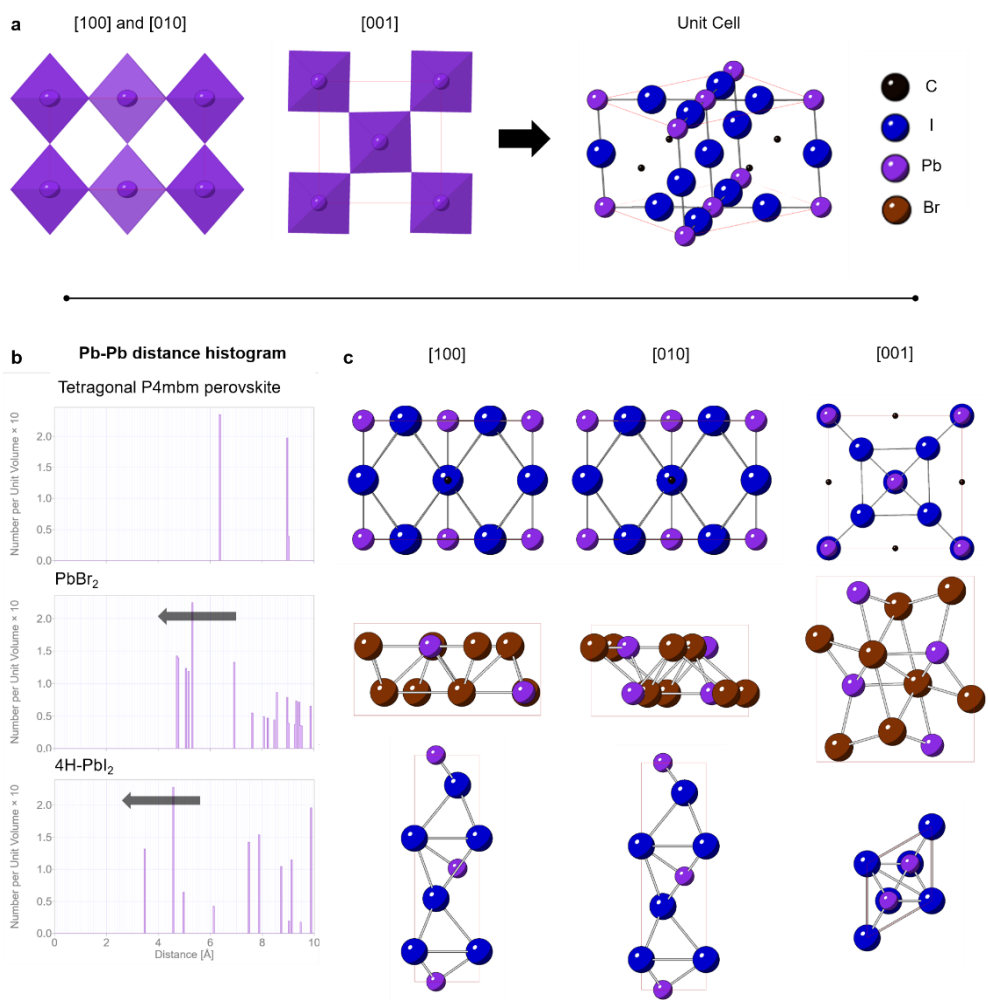

*Figure S9- Schematics of some of the crystal structure files used for the simulation and indexing of experimental diffraction patterns. (a) Schematic of how the tetragonal P4mbm perovskite unit cell is built from the $BX_6$ octahedra. b) Histograms and c) the unit cells at the a, b and c directions for the perovskite, the $PbBr_2$ and the $4H-PbI_2$ phases.*

S12

# Discussion on the direct beam intensity

The datasets reported in this work were acquired continuously over the course of more than 3 hours. A non-linear darkening of the direct beam over the series of scans is observed, with the direct beam intensity of the final scan being 6x lower intensity than the initial intensity (see Figure S10a-b). Such effect could be explained by a series of factor:

Changes in the emission values of the electron gun are unlikely to account for such large drop of intensity. Performance tests after flashing the electron emission gun at 300 keV in STEM mode were completed at ePSIC. While the emission current, measured after the gun accelerators and before the condenser apertures, reduced by ~20% after 3 hours, the probe current, measured using a Faraday cup after the whole column, exhibited stable constant beam currents until after 7 hours (see Figure S10c). Moreover, other SED data was acquired under the same beam conditions after the last frame reported in this work, with direct beam currents saturating the central pixels of the camera. These findings suggest the emission current of the electron gun did not change significantly during the acquisition time.

Saturation at the direct beam position can occur due to the bit-depth limitations of the Medipix chip when operating at readout times (1 ms) necessary for accessing low electron fluence at the sample. At the initial scanned frame, some diffraction patterns exhibit saturation at the direct beam and some of the diffracted spots. Over continuous scanning, the detected central beam intensity decreases, no longer exhibiting saturation at the direct beam. With the creation of pinholes in the sample, some areas in the scanned region would be expected to show saturation (0-63 counts per pixel at 64-bit depth). Nevertheless, this is not observed, so saturation does not affect the diffracted reflections in this study. We can obtain the time evolution of the direct beam intensity and shape profile from a region from which no saturation is observed at any of the frames. The direct beam intensity decay can be seen in Figure S10.

Sample thickening due to deposition of carbon contamination can increase sample thickness and thus reduce the detected direct beam intensity. To estimate the thickness of amorphous carbon required to attenuate the electron direct beam by ~50% of the initial intensity, the Poison inelastic scattering model can be used as described below (see Eq. S2 and S3). Under the experimental conditions used in this work and with a $Z_{eff}$ of 6 atomic units, a conservative thickness approximation of ~190 nm would be required for P0 to be reduced by 50%. These estimations use the over-simplistic Poison inelastic model and should be taken as qualitative explanations only, suggesting that carbon contamination is significant.

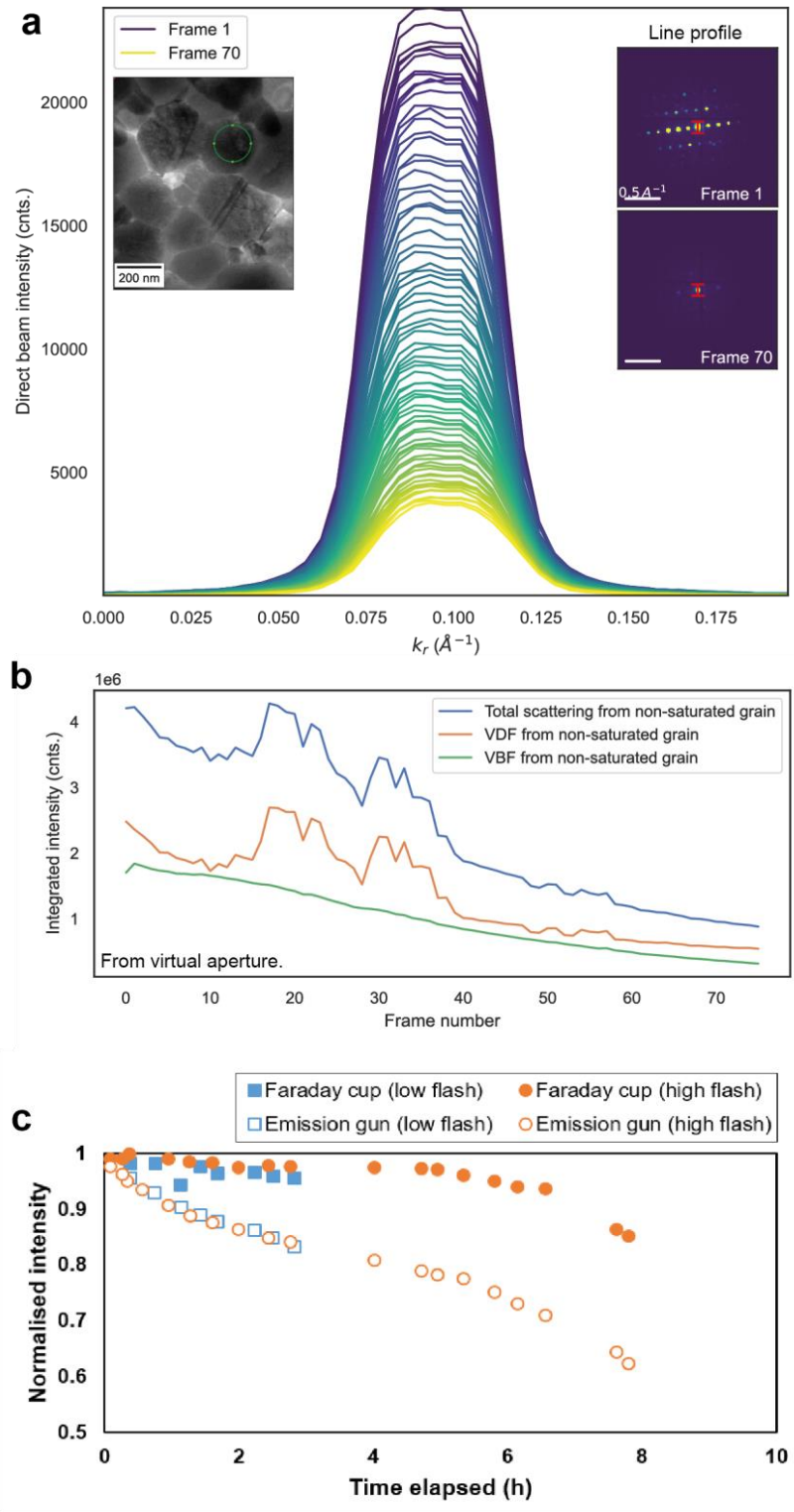

*Figure S10- a) Line profile of the direct beam over exposure evolution. Inset on the left shows the grain from which diffraction was taken. Inset on the right shows the line from which the intensity profile are extracted from. b) Integrated intensity evolution over frame number taken from integrating the diffraction intensity using the virtual apertures shown in Figure S2.. c) Emission stability performance test of the emission electron gun over 7 hours at 300 kV (acceleration 1 3.49 kV, acceleration 2 5.00 kV, initial emission 16.1 µA) after high and low flashing in STEM mode (9 cm camera length, CL aperture 2 (30 µm) and IL3 changed to 835F in FLC mode to make a small beam to fit the Faraday cup). The emission value drops due to the beam profile changing at higher angles, which is not detected down-stream on Faraday cup. Data acquired by Mohsen Danaie (ePSIC).*

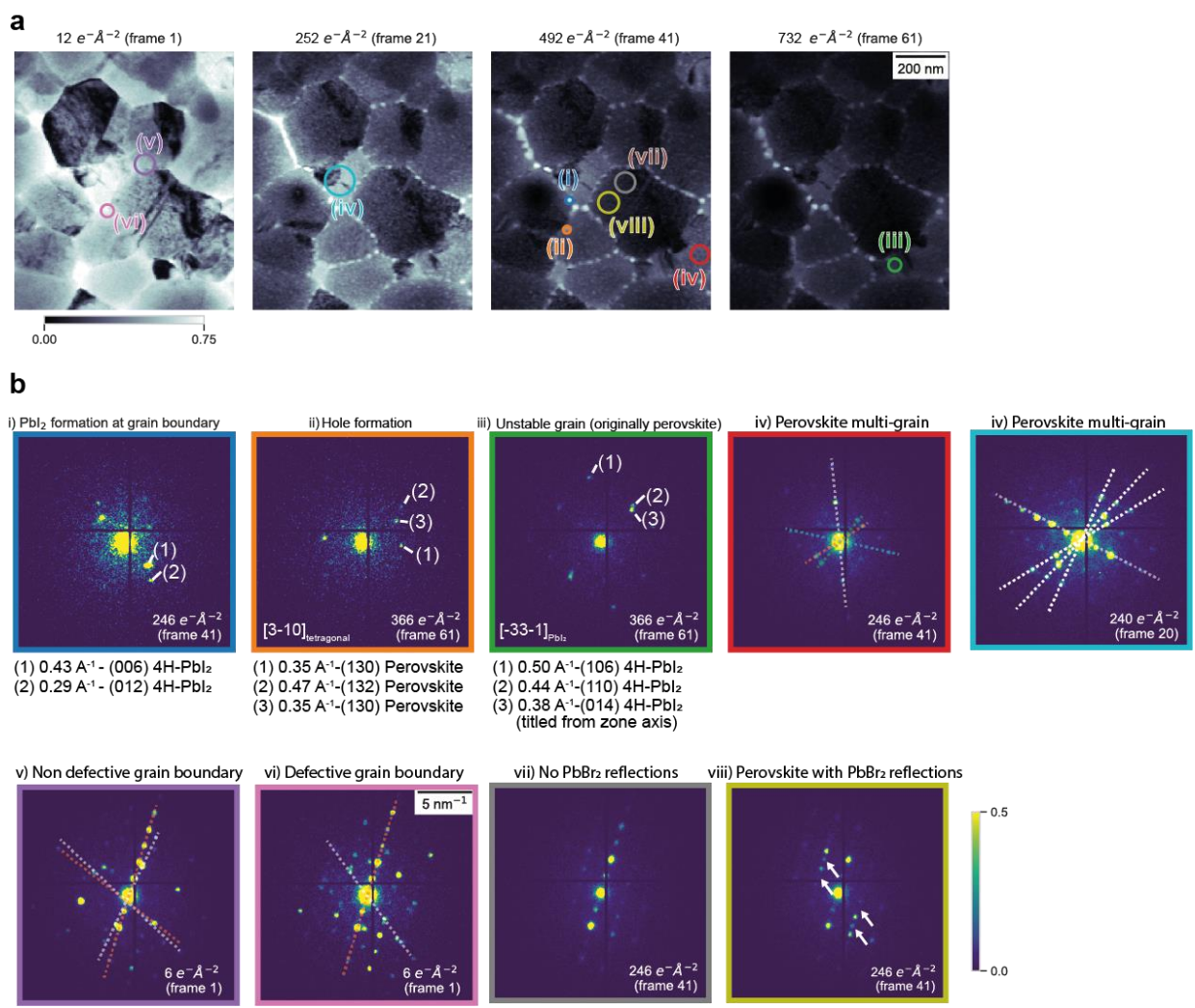

*Figure S11- Supporting figure to Fig. 3 in the main text. Additional examples of (blue and green) PbI₂ formation at the grain boundary, (orange) hole formation, showing diffuse scattering and absence of Bragg peaks other than the ones from the adjacent grains, (red) a series of small grains indexable to the tetragonal P4mbm perovskite structure but at different crystal rotations (similar to the perovskite multi-grain identified in Fig. 3 in the main text). In (purple) and (pink) an additional example of the initial state of two adjacent grain boundaries across the large perovskite grain showing a clean and a defective boundary, respectively. An additional example of a tetragonal perovskite grain showing additional PbBr₂ reflections appearing heterogeneously across the grain is shown in (grey) and (gold).*

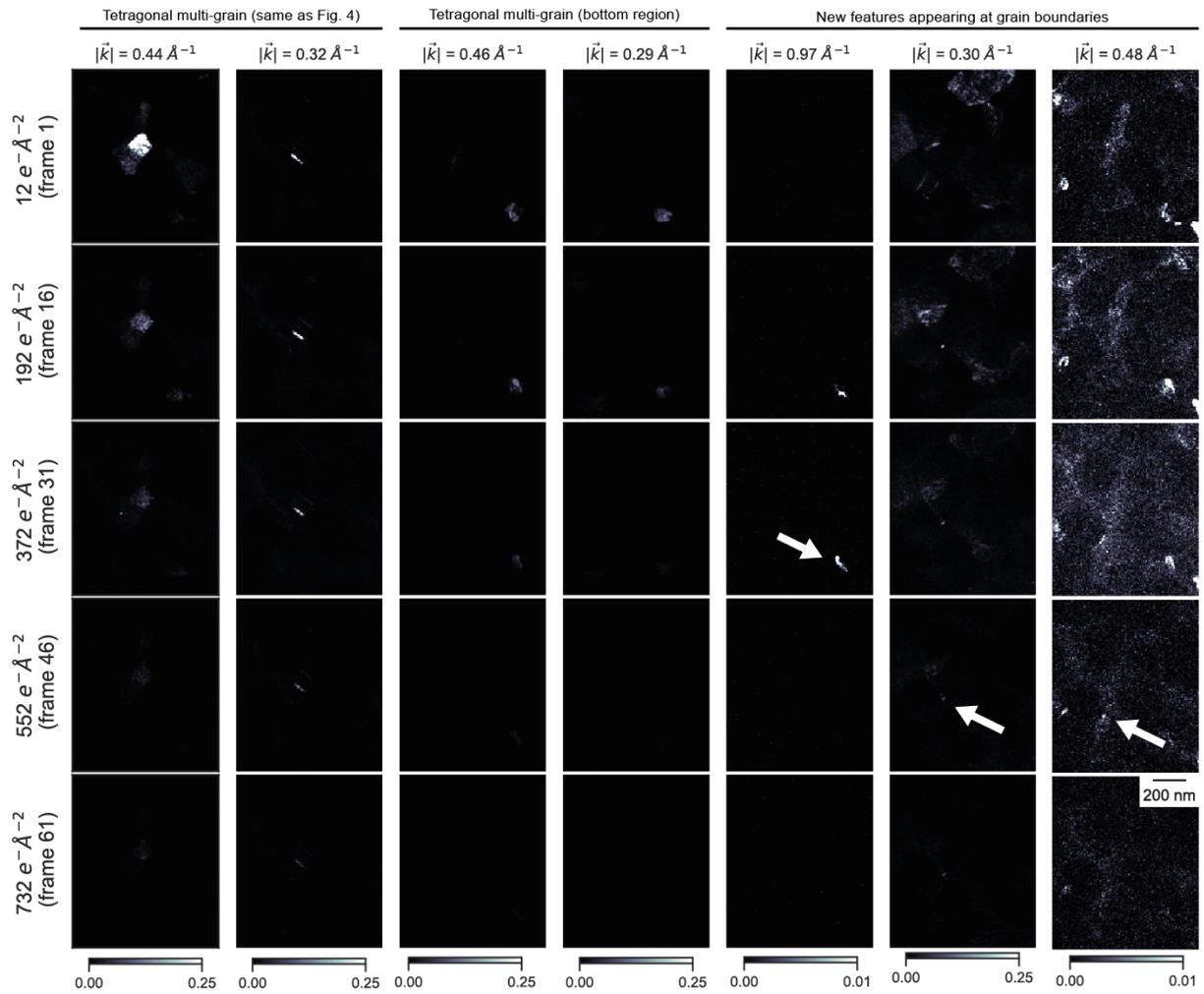

*Figure S12- Supporting figure to Fig. 4 in the main text, showing the evolution of vDF images as a function of electron exposure. Additional examples for tetragonal perovskite peaks and additional reflections appearing after the first scan.*

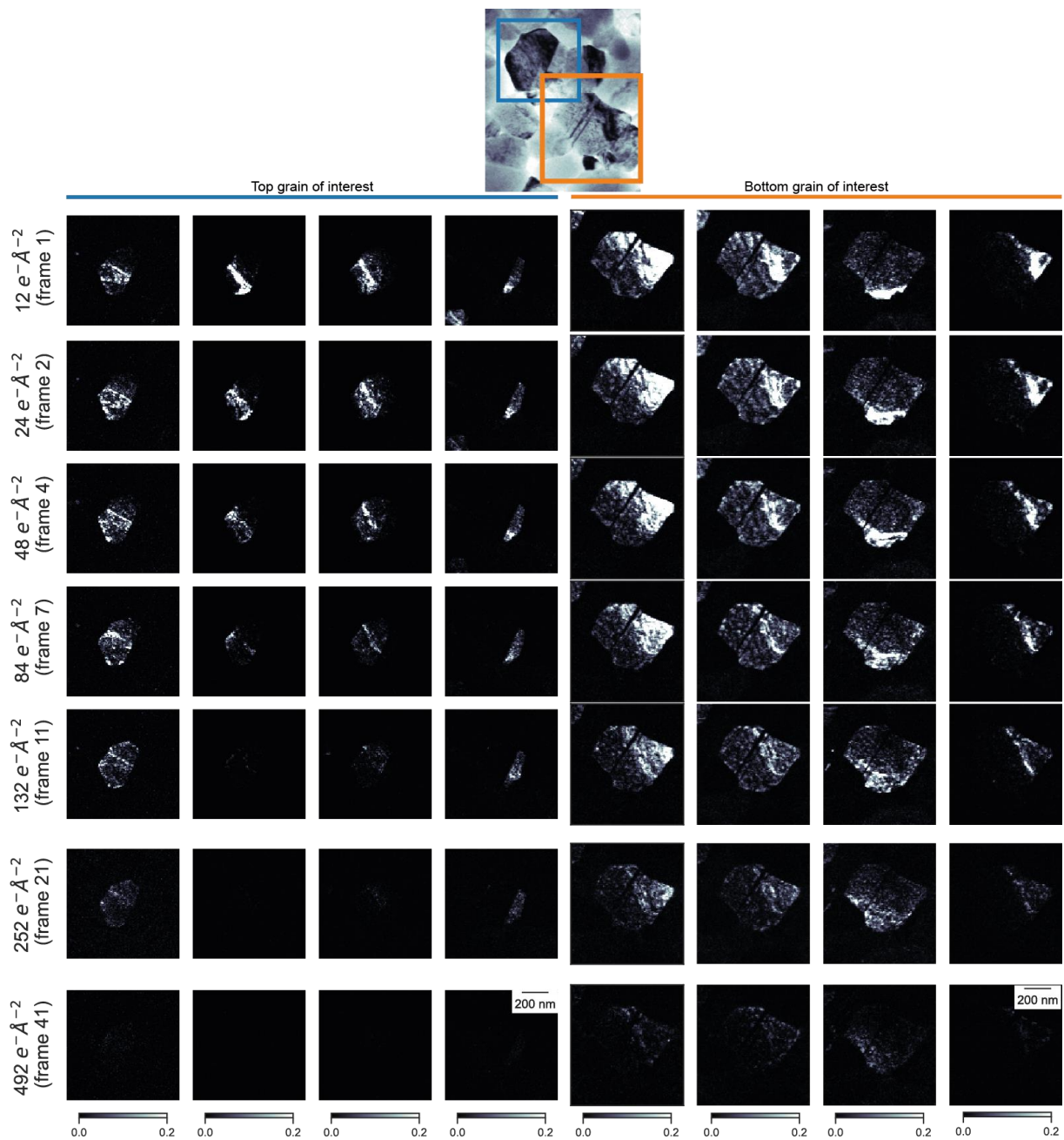

*Figure S13- vDF images of the grain tilting observed within the 2 main grains of interest and the progressive amorphisation of the perovskite grain. Such rotations are also seen in the adjacent grains.*

# Degradation effect on the crystallographic octahedral tilt

A grain oriented down near the [001] zone axis is found at a region far away from the centre where beam parking is located. Figure S14a shows the location of the grain, its mean diffraction pattern and the simulation of the [001] zone axis. At this zone axis, superstructure reflections are visible (see orange circle), attributed to the corner-sharing $BX_6$ octahedron tilt from the tetragonal perovskite phase. These superstructure reflections eventually disappear, as the sample is exposed to higher fluence and damaged (see white arrows in Figure S14b-c). The faster loss of superstructure reflections in comparison to the loss of the main bright reflections may suggest the loss of the tetragonal distortion trending towards a more cubic structure. Since the tetragonal distortion arises from cation alloying,[10] the loss of the more volatile MA and or FA cations is likely to tend towards a cubic structure. These results are consistent with the reported degradation pathway described in this work. These results also offer more evidence that such degradation pathway is also observed at regions away from where beam parking is happening.

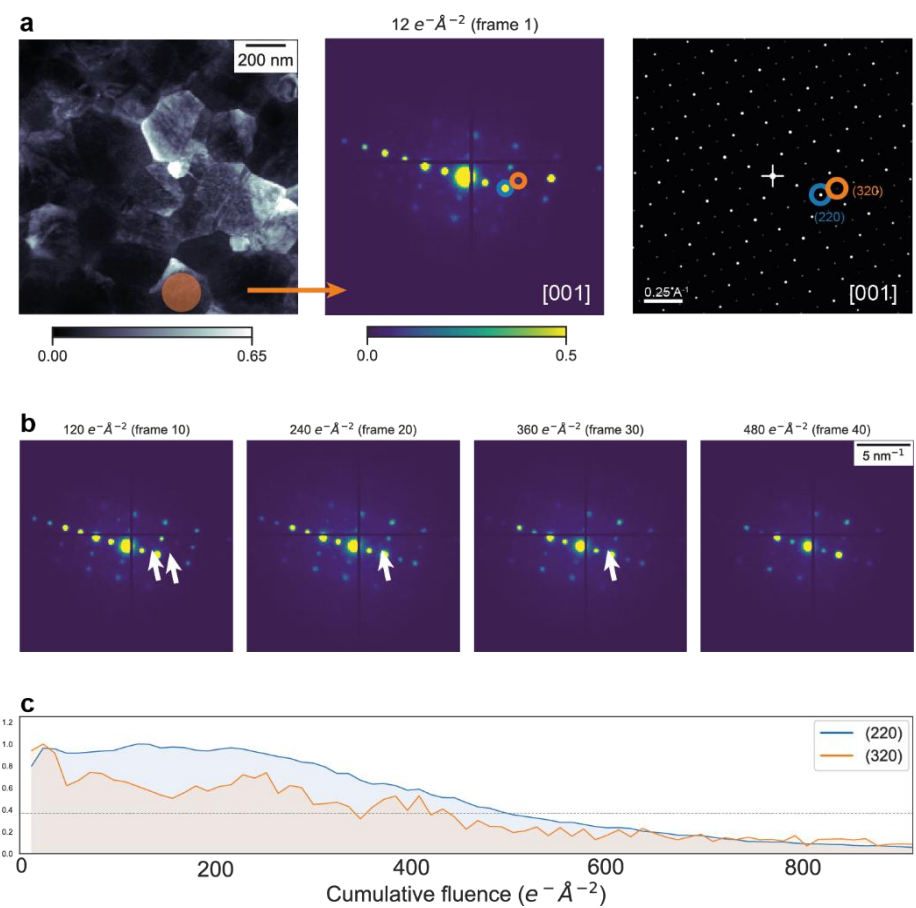

*Figure S14- SED evolution of a grain at the [001] zone axis of a halide perovskite film. a) The grain location is away from the beam-parking position, as shown in the vBF image at the 1st frame (left). The mean diffraction pattern from the orange circle region is shown at the centre and the simulated diffraction pattern at the [001] zone axis at the right, especially showing weak reflections in orange. b) The evolution of the mean diffraction pattern over increased fluence. White arrows show the disappearance of the weak additional reflections due to octahedral tilt. d) Comparison between intensity profiles taken from the (220) and the (320) tetragonal diffraction spots.*

# Supplementary X-ray diffraction data and simulation

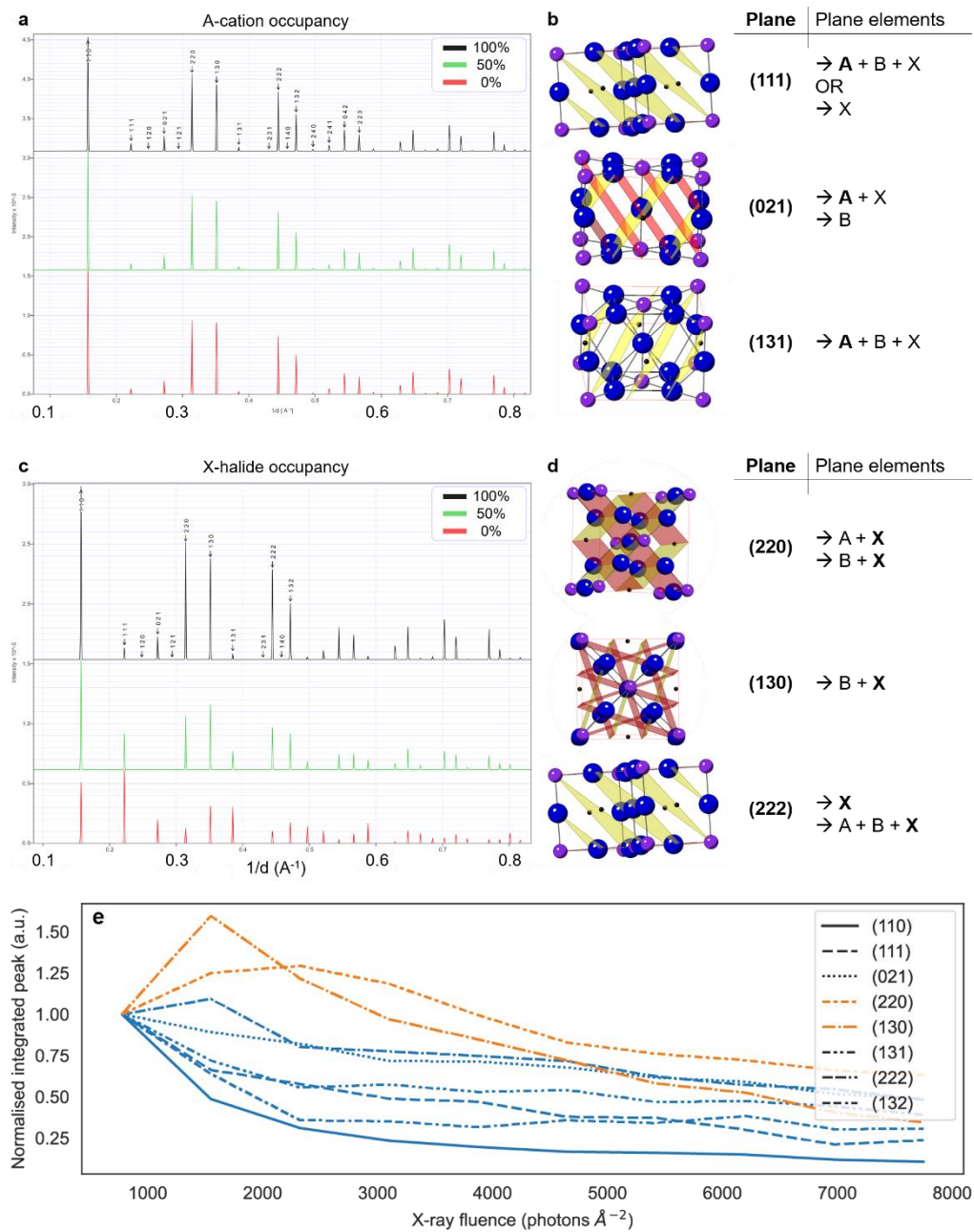

*Figure S15- Simulation of the effect of the occupation of a) the A-cation (FA) and c) the X-halide anion (I) on the 1D diffraction patterns. On the left, schematics of the specific crystallographic planes which are most affected by the change of occupancy in b) A-cation and d) the X-halide, and how they relate to the elements on those respective planes. e) Degradation rate of the integrated peaks for the nXRD mean diffraction patterns in Fig. 5 in the main text. In orange the slowest decaying crystal planes.*

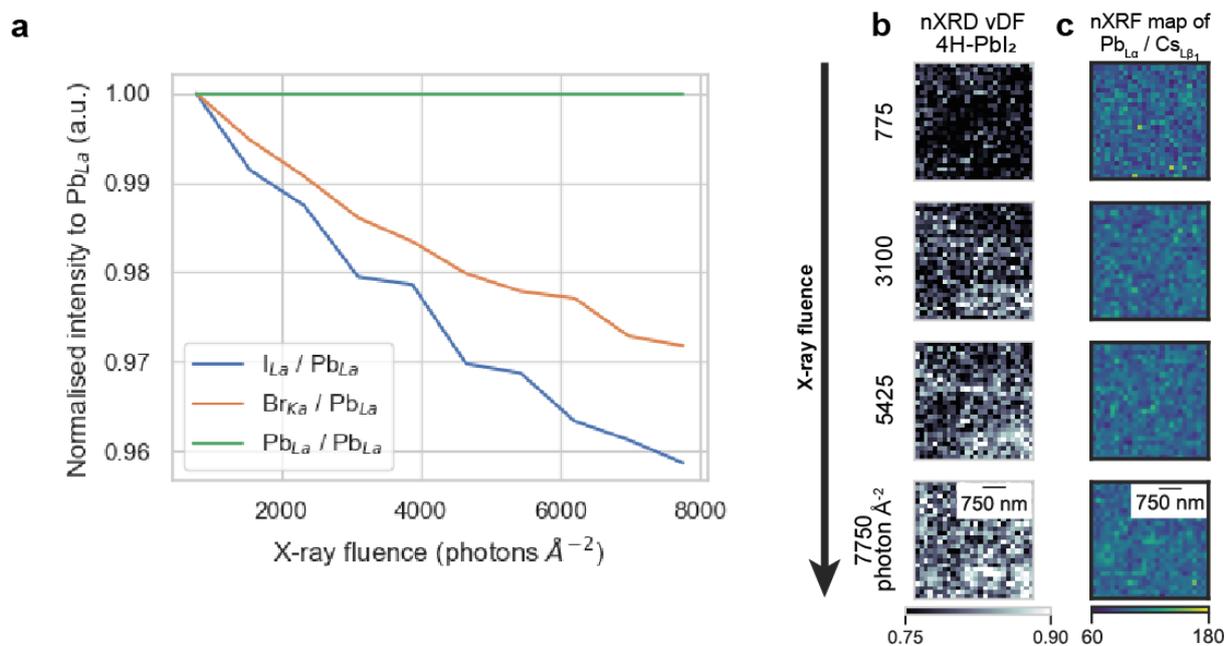

*Figure S16- Supplementary nXRD data. a) nXRF summed intensity over the scanned area against X-ray fluence for the Pb and halides. b) vDF images taken from solely the intensity of the appearing peak (\*) shown in Fig. 5 in the main text, showing the appearance of such peak homogeneously across the whole scanned area. c) nXRF map of the Cs distribution over X-ray fluence (normalised by the $Pb_{La}$ peak). No Cs segregation is observed. The peak $Cs_{Lb1}$ was selected, as the brighter $Cs_{La}$ overlaps with the $I_{Lb1}$ peak.*

**Additional synchrotron X-ray diffraction experiment (staticXRD):** *Calibration of fluence.* The photon flux in the static box-beam XRD experiment was indirectly measured by the current passing through an ion chamber during each image acquisition. The current flux fluctuated between $3.8 - 4.5 \times 10^{-13}$ $A$, at a mean value of $4.3 \times 10^{-13}$ $A$. These measurements are electrons detected from the ionization of the gas in the chamber. Five basic interactions can occur when X-rays penetrate a material, the main ones being photoelectric, Compton and Rayleigh scattering. In order to get the flux in photons per second, an estimation of the flux required to create the readout electron ion pair formation in $N_2$ gas flow (within the ionization chamber). The tabulated cross sections for photoelectric, Compton and Rayleigh effects on the specific $N_2$ gas and 17.2 keV beam energy are: $c_{rayleigh}$, $c_{compton}$, $c_{photo}$ = 2.3634, 3.5412, 15.0215 respectively. Assuming the photoelectric effect is the dominant one (approximately true for beam energies below 30 keV),[11] all the electron ion pairs measured in the gas chamber are only produced by the photoelectric effect, which is 71.78 % of the total scattering cross sections. These assumptions estimate a flux fluctuation between $6.3 - 7.5 \times 10^{11}$ $photons\ s^{-1}$, at a mean flux of $7.2 \times 10^{11}$ $photons\ s^{-1}$. Fluence (flux density) can be estimated from the X-ray beam size of ~1 μm$^2$ and the dwell time per frame of 1.64 s, yielding a mean fluence of 12,000 photons Å$^{-2}$ for each frame.

Figure S17 shows the radially integrated static XRD evolution of a halide perovskite film, of same composition as reported in the text, over accumulated illumination for this static beam experiment. The peak degradation follows similar patterns as seen in Fig. 5a in the main text, yet degradation of the main peaks is only discernible at 2-3 order of magnitude higher fluence than for the nXRD experiment, due to the larger size of the X-ray beam.

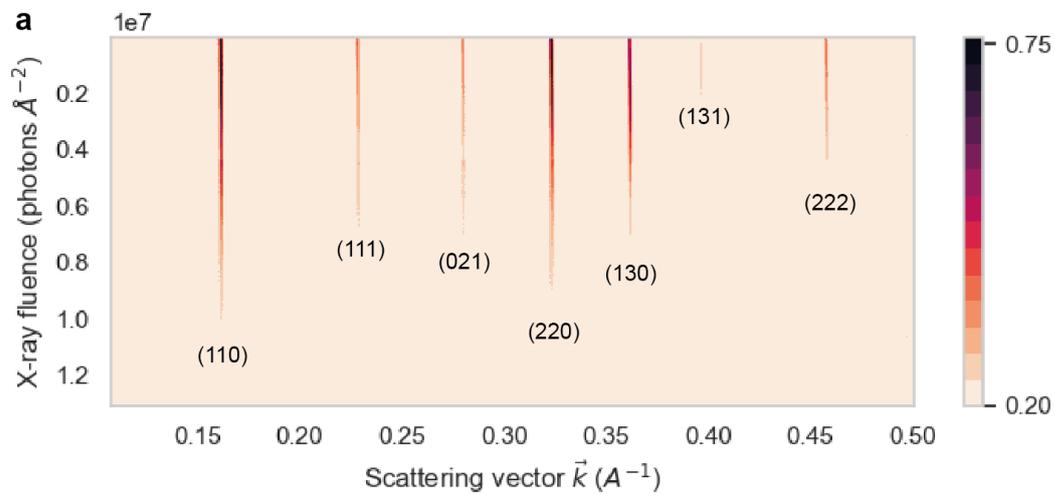

*Figure S17- Radially integrated static XRD evolution of a halide perovskite film over accumulated illumination indexed to the tetragonal perovskite phase. The calibration of the scattering vector is within an error of +- 0.05 A. The peak degradation follows similar patterns as seen in Fig. 5a in the main text but at 2-3 order of magnitude higher fluence due to the larger size of the X-ray beam.*

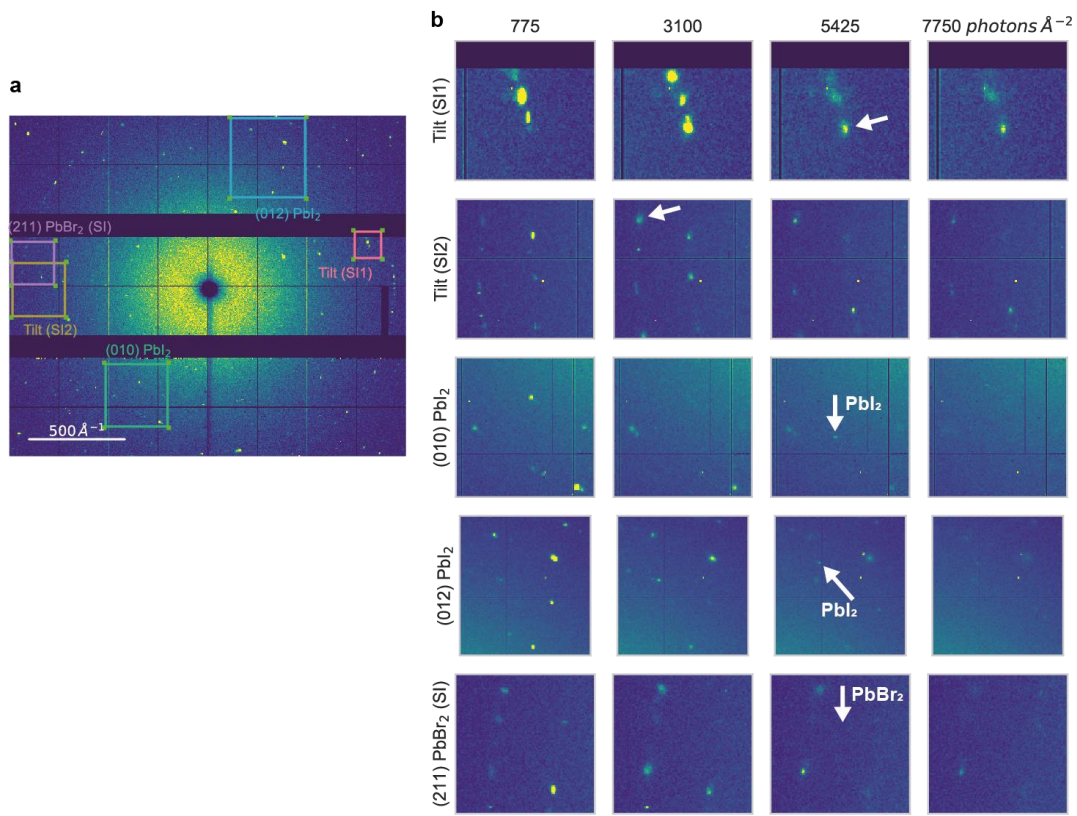

*Figure S18- Supplementary figure to Fig. 5 in the main text. a) Mean 2D nXRD detector image, with some areas of interest. b) Evolution of the features seen in the raw 2D detector over fluence, showing crystal tilting and the appearance of additional reflections indexable to both $PbI_2$ and $PbBr_2$.*

# Discussion on elastic and inelastic scattering

**Poisson model for inelastic scattering of electrons:** Electrons are known to scatter inelastically with increasing probability for increasing sample thickness. In this study, electron transparent samples of ~200 nm were used. This thickness can lead to significant inelastic scattering. An estimation of the inelastic mean free path on the basis of the dipole formula can be found from the definitions from Egerton:[12]

$$\lambda = \frac{106 F \left(\frac{E_0}{E_m}\right)}{\ln\left(\frac{2\beta E_0}{E_m}\right)} \quad (S2)$$

With F: tabulated relativistic factor at 200 keV, $E_0$: the bandgap, $\beta$: the convergence angle, and $E_m$: the mean energy loss energy, which can be estimated using $E_m \sim 7.6\, Z_{eff}^{0.36}$. For the $(FA_{0.79}MA_{0.16}Cs_{0.05})Pb(I_{0.83}Br_{0.17})_3$ perovskite composition, $Z_{eff}$ is 36.4 atomic units. For $\beta = 1$ mrad at 200 keV, a mean free path $\lambda$ is found of 177 nm.

At $t = 200$ nm, the possibility of inelastic scattering can be estimated using Poisson's law:

$$P_n = \left(\frac{1}{n!}\right)\left(\frac{t}{\lambda}\right)^n \exp\left(-\frac{t}{\lambda}\right) \quad (S3)$$

Where n is the number of scattering events. These estimations yield probabilities of:

*Table S1 – Probability of no, single and multiple inelastic scattering.*

| | |
|---|---|
| No inelastic scattering | P0 = 32% |
| Single scattering | P1 = 37% |
| Double scattering | P2 = 21% |
| Triple scattering | P3 = 8% |

Therefore, from the moment the electrons interact with the sample and from all elastic scattering events, a significant percentage of electrons can scatter inelastically multiple times transferring energy to the sample.

**Elastic and inelastic cross section ratio for electrons:** A derivation of the estimation of the inelastic to elastic mean free path can be found from the definitions from Egerton.[12] The following parameters were approximated: collection angle $b$ =10 mrad (does not affect the ratio between the cross sections), and the following sample specific parameters were used: $Z_{eff}$ is 36.4 atomic units, mean energy loss 27.2 eV, and atomic density of ~7 nm$^{-3}$.

Figure S19a shows the log-log plot of the ratio between the inelastic to elastic scattering cross sections for electrons against the atomic number. The relation is linear in the log-log scale, not affected substantially by acceleration voltage from 80 to 200 keV. From these approximations a relationship between the cross section ratios ($\frac{\sigma_i}{\sigma_e}$) and the atomic number ($Z$) is found:

$$Electrons \rightarrow \frac{\sigma_i}{\sigma_e} \propto 19 \cdot Z^{-1} \quad (S4)$$

**Elastic and inelastic cross section ratio for X-rays:** The elastic and inelastic cross sections for high-energy photons are retrieved from the XCOM Photon Cross Sections Database.[13] This web database provides calculated photon cross sections for scattering cross sections, as well as photoelectric absorption, pair production, total attenuation coefficients. The database includes any element ($Z \leq 100$) at energies from 1 keV to 100 GeV. A custom-made script was used to retrieve all relevant data at the relevant energies used in this study. Coherent cross sections were used as elastic scattering cross sections and incoherent as inelastic ones.

Figure S19b shows the log-log plot of the ratio between the incoherent to coherent (referred as inelastic to elastic) scattering cross sections for photons against the atomic number. The relation is linear in the log-log scale. From these values at 20 keV, a relationship between the cross section ratios ($\frac{\sigma_i}{\sigma_e}$) and the atomic number ($Z$) is found:

$$20 \; keV \; Photons \rightarrow \frac{\sigma_i}{\sigma_e} \propto 50 \cdot Z^{-1.7} \quad (S5)$$

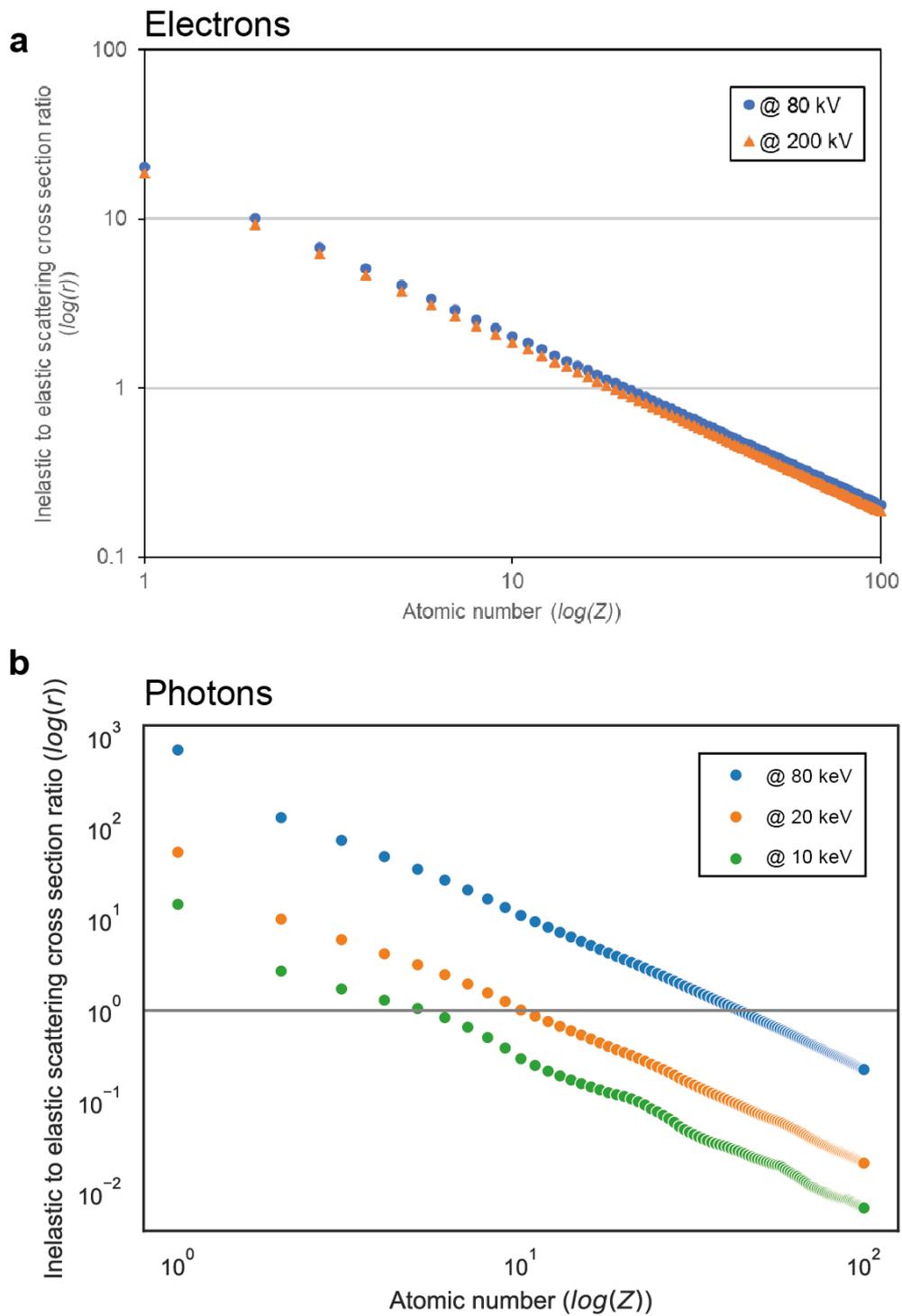

*Figure S19 – Inelastic to elastic scattering cross section ratios for a) electrons and b) photons at different energies. In blue, at the comparable energy of 80 keV. In orange, at the acceleration energies used for the data acquisition of this work.*

# Charge carrier concentration estimation

The illumination of any semiconductor with radiation can produce the excitation of charge carriers (electron and hole pairs) from the valence to the conduction band. These charge carriers can diffuse across the grains, of mean size around ~100 nm. The carriers decay after a certain lifetime, shown to be of the order of hundreds of nanoseconds from photoluminescence studies of similar compositions,[14] but much shorter lived (<1 nm) from comparable cathodoluminescence studies.[15] Moreover, high-energy beams tend to induce high densities of charge carriers, triggering Auger processes which result in even shorter lifetimes.[14] These findings suggest that the acquisition dwell time used here (1 ms and 0.5 s for SED and nXRD, respectively) is significantly longer than the charge carrier lifetimes, avoiding any accumulation of carriers from pixel-to-pixel scanned.

The charge carrier density produced by both the electron and the X-ray beam can be estimated as follows.

**SED:** Electron diffraction data was acquired at 200 keV and at 12 e$^-$ Å$^{-2}$ per frame. Since the electron beam has a significantly larger energy than the band gap of the material (~1.653 eV) and since measurements are taken in transmittance mode, only the mean energy-loss energy from inelastic scattering is considered capable of creating charge carriers. For the acquisition conditions of the SED experiments (200 keV, 1 mrad collection angle, effective atomic mass effective Z = 36.4, 200 nm sample thickness), this inelastic scattering mean energy loss is ~27.2 eV (refer to Equation S2: $\lambda = \frac{106 F\left(\frac{E_0}{E_m}\right)}{\ln\left(\frac{2\beta E_0}{E_m}\right)}$ (S2)). The inelastic scattering probability of single, double and triple scattering is of ~66%. Therefore, making the simplistic assumption that only 8 e$^-$ Å$^{-2}$ of the incoming fluence will scatter inelastically, each electron can approximately excite $\sim E_0 / 3E_g \approx 6$ e$^-$–h$^+$ pairs.[16] To estimate the lower and upper bounds of the charge carrier density, we assume a diffusion area ranging from the size of the beam (~5 nm diameter) up to the size of the mean grain size (~100 nm). Assuming also a mean interaction depth of 200 nm, due to the thin nature of the specimen for TEM studies, an interaction volume of $(\frac{5\ to\ 100}{2})^2 \cdot \pi\ nm^2 \times 200\ nm = 10^{-15}\ to\ 10^{-18} cm^3$ is estimated for each electron. Therefore, the charge carrier density of the localised excitation is $3 \times 10^{15}\ to\ 1 \times 10^{18}$ e$^-$–h$^+$-pairs cm$^{-3}$ per incoming electron. This translates to a constant local excitation of $\mathbf{10^{16}\ to\ 10^{19}}$ **e$^-$–h$^+$ pairs cm$^{-3}$** per frame.

**nXRD:** Diffraction data was acquired at 20 keV and at 775 photons Å$^{-2}$, as calibrated in the previous section in the SI. In order to estimate the material transmittance at 20 keV, the pyMca 5.5.4 tool was used,[17] assuming a density of 4.16 g cm$^{-3}$,[18] a film thickness of 600 nm and the film composition described in the methods. With an estimated transmittance coefficient of μ = 0.984, a total fluence of 12.4 photons Å$^{-2}$ will interact with the sample. We assume as an overestimation that each of the 20 keV photon can produce $\sim E_0 / E_g \approx 10^4$ e$^-$–h$^+$ pairs. To estimate the lower and upper bounds of the charge carrier density, we assume a diffusion area ranging from the size of the beam (~150 nm diameter) down to the size of the mean grain size (~100 nm). From the interaction of 600 nm, the thickness of the specimen, an interaction volume of $(\frac{100\ to\ 150}{2})^2 \cdot \pi\ nm^2 \times 600\ nm = 10^{-14}\ to\ 10^{-15} cm^3$ is estimated for each photon, resulting in an overestimated localised excitation of $1\ to\ 3 \times 10^{18}$ e$^-$–h$^+$-pairs cm$^{-3}$ per photon, or **$10^{19}$ e$^-$–h$^+$ pairs cm$^{-3}$** per frame.

Again, these amount of charge carrier densities should be taken as over-simplistic approximations. However, the charge carrier densities from the two different techniques are in a similar order of magnitude, higher than the charge carrier densities of these materials under 1 sun illumination ($10^{14}\ to\ 10^{16}$ e$^-$–h$^+$-pairs cm$^{-3}$).

## Simulation of electron radiation in orbit

The electron spectra from the safer orbit near the planet Earth, and the harsher orbit in Jupiter were simulated using SPENVIS.[19] The fluence of the electron spectrum at 200 keV energy for both orbits was used as input to relate the fluence thresholds found in this study to space applications, as shown in Figure S20. The fluxes were calculated from the default orbits generated by SPENVIS for each planet and the default models for electrons for each planet (AE-8 MAX and D&G83 for the Earth and Jupiter, respectively). With a total flux of about $3 \times 10^{-9}$ e$^-$ Å$^{-2}$ s$^{-1}$ at 200 keV in the Earth orbit, a fluence of 200 e$^-$ Å$^{-2}$ would be accumulated after around $7 \times 10^{10}\ s \approx 2000$ years. The same fluence threshold would be reached earlier at the harsher Jupiter orbit after only ~200 years (with a total flux of ~$10^{-8}$ e$^-$ Å$^{-2}$ s$^{-1}$). Despite the simplicity of these estimates, the resilience of these materials to degradation from electron radiation is promising for space PV applications as the radiation fluence used in microscopy are significantly larger than the radiation in space by several orders of magnitude.

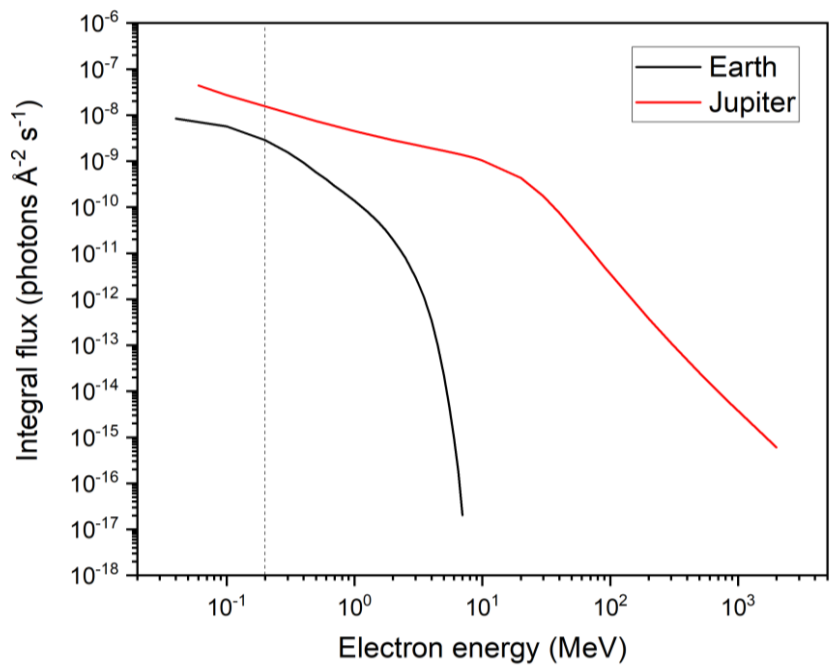

*Figure S20- Fluence of the electron spectrum radiation in orbits in the Earth (black) and in Jupiter (red), as simulated using SPENVIS.[19] This spectrum was used to relate the fluence from the experimental results of this work to space PV applications. As a dashed vertical line, the acceleration voltage used for the SED acquisition.*